\documentclass[structabstract]{aa}
\usepackage{graphicx}
\usepackage{natbib}
\newcommand{\Msun}{\ensuremath{\,{M}_\odot}}                      % Solar mass symbol
\newcommand{\Rsun}{\ensuremath{\,{R}_\odot}}                      % Solar radius symbol
\newcommand{\Mjup}{\ensuremath{\,{M}_{\rm Jup}}}                  % Jupiter mass symbol
\newcommand{\Rjup}{\ensuremath{\,{R}_{\rm Jup}}}                  % Jupiter radius symbol
\newcommand{\safronov}{\ensuremath{\Theta}}                       % Safronov number symbol
\newcommand{\kms}{\,km\,s$^{-1}$}                                 % km/s symbol
\newcommand{\mss}{\,m\,s$^{-2}$}                                  % m/s^2 symbol
\newcommand{\pjup}{\ensuremath{\,\rho_{\rm Jup}}}                 % Jupiter density symbol
\newcommand{\psun}{\ensuremath{\,\rho_\odot}}                     % Solar density symbol
\newcommand{\mc}[1]{\multicolumn{2}{c}{#1}}

\newcommand{\erc}[3]{\mc{\ensuremath{#1^{+#2}_{-#3}}}}

\usepackage{amsmath,amssymb,graphicx}
\begin{document}
%
%\linenumbers

\title{The GAPS Programme with HARPS-N at TNG\thanks{Based on observations made with ($i$) the
Italian 3.58\,m Telescopio Nazionale Galileo at the Observatory of
Roque de los Muchachos, ($ii$) the Cassini 1.52\,m telescope at
the Astronomical Observatory of Bologna, ($iii$) the Zeiss 1.23\,m
telescope at the Observatory of Calar Alto, and the IAC 80\,cm
telescope at the Teide Observatory.}}
   \subtitle{
VIII: Observations of the Rossiter-McLaughlin effect and
characterisation of the transiting planetary systems HAT-P-36 and
WASP-11/HAT-P-10}
\titlerunning{Physical properties of HAT-P-36\,b \& WASP-11/HAT-P-10\,b}

   \author{
          L. Mancini \inst{1,2}
          \and
          M. Esposito\inst{3,4}
          \and
          E. Covino\inst{5}
          \and
%%%%%%%%%%%%%%%%%%%%%%%%%%%%%%%%%%%%%%%%%%%
          G. Raia\inst{5}
          \and
          J. Southworth\inst{6}
          \and
          J. Tregloan-Reed\inst{7}
          \and
%%%%%%%%%%%%%%%%%%%%%%%%%%%%%%%%%%%%%%%%%%%
          K. Biazzo\inst{8}
          \and
          A. Bonomo\inst{2}
          \and
          S. Desidera\inst{9}
          \and
          A.~F. Lanza\inst{8}
          \and
          G. Maciejewski\inst{10}
          \and
          E. Poretti\inst{11}
          \and
          A. Sozzetti\inst{2}
          \and
%%%%%%%%%%%%%%%%%%%%%%%%%%%%%%%%%%%%%%%%%%%
          F. Borsa\inst{11}
          \and
          I. Bruni\inst{12}
          \and
          S. Ciceri\inst{1}
          \and
          R. Claudi\inst{9}
          \and
          R. Cosentino\inst{13}
          \and
          R. Gratton\inst{9}
          \and
          A.~F. Martinez Fiorenzano\inst{13}
          \and
          G. Lodato\inst{14}
          \and
          V. Lorenzi\inst{13}
          \and
          F. Marzari\inst{9,13}
          \and
          S. Murabito\inst{3,4}
          \and
%%%%%%%%%%%%%%%%%%%%%%%%%%%%%%%%%%%%%%%%%%%
          L. Affer\inst{15}
          \and
         A. Bignamini\inst{16}
          \and
         L.~R. Bedin\inst{9}
          \and
          C. Boccato\inst{9}
          \and
          M. Damasso\inst{2}
          \and
          Th. Henning\inst{1}
          \and
          A. Maggio\inst{15}
          \and
          G. Micela\inst{15}
          \and
          E. Molinari\inst{13,17}
          \and
          I. Pagano\inst{8}
          \and
          G. Piotto\inst{9,18}
          \and
          M. Rainer\inst{11}
          \and
          G. Scandariato\inst{8}
          \and
          R. Smareglia\inst{16}
          \and
          R. Zanmar Sanchez\inst{8}}
{
    % 1
    \institute{Max Planck Institute for Astronomy, K\"{o}nigstuhl 17, 69117 -- Heidelberg, Germany \\
    \email{mancini@mpia.de}
    \and
    % 2
    INAF -- Osservatorio Astrofisico di Torino, via Osservatorio 20, 10025 -- Pino Torinese, Italy
    \and
    % 3
    Instituto de Astrof\'{i}sica de Canarias, C/ V\'{i}a L\'{a}ctea s/n, 38205 -- La Laguna, Tenerife, Spain
    \and
    % 4
    Dep. de Astrof\'{i}sica, Universidad de La Laguna, Avda. Astrof\'{i}sico F. S\'{a}nchez s/n, 38206 La Laguna, Tenerife, Spain
    \and
    % 5
    INAF -- Osservatorio Astronomico di Capodimonte, via Moiariello 16, 80131 -- Naples, Italy
    \and
    % 6
    Astrophysics Group, Keele University, Keele ST5 5BG, UK
    \and
    % 7
    NASA Ames Research Center, Moffett Field, CA 94035, USA
    \and
    % 8
    INAF -- Osservatorio Astrofisico di Catania, via S. Sofia 78, 95123 -- Catania, Italy
    \and
    % 9
    INAF -- Osservatorio Astronomico di Padova, Vicolo dell'Osservatorio 5, 35122 -- Padova, Italy
    \and
    % 10
    Centre for Astronomy, Nicolaus Copernicus University, Grudziadzka 5, 87-100 -- Torun, Poland
    \and
    % 11
    INAF -- Osservatorio Astronomico di Brera, via E. Bianchi 46, 23807 -- Merate (Lecco), Italy
    \and
    % 12
    INAF -- Osservatorio Astronomico di Bologna, Via Ranzani 1, 40127 -- Bologna, Italy
    \and
    % 13
    Fundaci\'{o}n Galileo Galilei - INAF, Rambla Jos\'{e} Ana Fernandez P\'{e}rez, 738712 -- Bre\~{n}a Baja, Tenerife, Spain
    \and
    % 14
    Dipartimento di Fisica, Universit\`{a} di Milano, via Celoria 16, 20133 -- Milano, Italy
    \and
    % 15
    INAF -- Osservatorio Astronomico di Palermo, Piazza del Parlamento, 90134 -- Palermo, Italy
    \and
    % 16
    INAF -- Osservatorio Astronomico di Trieste, via Tiepolo 11, 34143 -- Trieste, Italy
    \and
    % 17
    INAF -- IASF Milano, via Bassini 15, 20133 -- Milano, Italy
    \and
    % 18
    Dip. di Fisica e Astronomia Galileo Galilei, Universit\`{a} di Padova, Vicolo dell'Osservatorio 2, 35122 -- Padova, Italy
    }

%   \date{Received ; Accepted}
 \abstract
 % 5 {} token are mandatory
{Orbital obliquity is thought to be a fundamental parameter in
tracing the physical mechanisms that cause the migration of giant
planets from the snow line down to roughly $10^{-2}$\,au from
their host stars. We are carrying out a large programme to
estimate the spin-orbit alignment of a sample of transiting
planetary systems to study what the possible configurations of
orbital obliquity are and wether they correlate with other stellar or
planetary properties.}
{We determine the true and the projected obliquity of
HAT-P-36 and WASP-11/HAT-P-10 systems, respectively, which are
both composed of a relatively cool star (with effective
temperature $T_{\mathrm{eff}}<6100$\, K) and a hot-Jupiter
planet.}
{Thanks to the high-resolution spectrograph HARPS-N, we observed
the Rossiter-McLaughlin effect for both the systems by acquiring
precise ($3-8$\,m\,s$^{-1}$) radial-velocity measurements during
planetary transit events. We also present photometric observations
comprising six light curves covering five transit events, which were obtained
using three medium-class telescopes. One transit of
WASP-11/HAT-P-10 was followed contemporaneously from two
observatories. The three transit light curves of HAT-P-36\,b show
anomalies attributable to starspot complexes on the surface of the
parent star, in agreement with the analysis of its spectra that
indicate a moderate activity
($\log{R_{\mathrm{HK}}^{\prime}}=-4.65$\,dex). By analysing the
complete HATNet data set of HAT-P-36, we estimated the stellar
rotation period by detecting a periodic photometric modulation in
the light curve caused by star spots, obtaining
$P_{\mathrm{rot}}=15.3 \pm 0.4$\,days, which implies that the
inclination of the stellar rotational axis with respect to the
line of sight is $i_{\star}=65^{\circ} \pm 34^{\circ}$.}
{We used the new spectroscopic and photometric data to revise the
main physical parameters and measure the sky-projected
misalignment angle of the two systems. We found $\lambda=
-14^{\circ} \pm 18^{\circ}$ for HAT-P-36 and $\lambda=7^{\circ}
\pm 5^{\circ}$ for WASP-11/HAT-P-10, indicating in both cases a
good spin-orbit alignment. In the case of HAT-P-36, we also
measured its real obliquity, which turned out to be
$\psi=25^{+38}_{-25}$\,degrees.}
{}
%{HAT-P-36 and WASP-11/HAT-P-10 join the large family of
%\textbf{(quite)} aligned planetary systems composed by cool stars
%and low-mass hot Jupiters. However, \textbf{the continuation of
%such observational programs is necessary as} more data are still
%needed to gain definitive understanding of the possible
%correlation between orbital obliquity and hot-Jupiter migration.}


\maketitle

% Sect. 1
%%%%%%%%%%%%%%%%%%%%%%%%%%%%%%%%%%%%%%%%%%%%%%%%%%%%%%
\section{Introduction}
\label{sec_1}
%%%%%%%%%%%%%%%%%%%%%%%%%%%%%%%%%%%%%%%%%%%%%%%%%%%%%%

Since its discovery \citep{mayor:1995}, hot Jupiters have
challenged the community of theoretical astrophysicists to explain
their existence. They are a population of gaseous giant extrasolar
planets, similar to Jupiter, but revolving very close to their
parent stars (between $\sim 0.01$ and $0.05$\,au), which causes
them to have orbital periods of few days and high equilibrium
temperatures (between $\sim 400$ and $2750$\,K)\footnote{Data
taken from TEPcat (Transiting Extrasolar Planet Catalogue),
available at www.astro.keele.ac.uk/jkt/tepcat/
\citep{southworth:2011}.}. According to the generally accepted
theory of planet formation (we refer the reader to the
review of \citealp{mordasini:2014}), hot Jupiters are thought to
form far from their parent stars, beyond the so-called \emph{snow
line}, and then \emph{migrate} inwards to their current positions
at a later time. However, the astrophysical mechanism that
regulates this migration process, as that which causes planets
moving on misaligned orbits, is still a matter of debate.

Spin-orbit obliquity, $\psi$, that is the angle between a planet's
orbital axis and its host star's spin, could be the primary tool
to understand the physics behind the migration process of giant
planets (e.g., \citealp{dawson:2014} and references therein). A
broad distribution of $\psi$ is expected for giant planets, whose
orbits are originally large (beyond the snow line) and (being
perturbed by other bodies) highly eccentric, and then become
smaller and circularize by planet's tidal dissipation during close
periastron passages. Instead, a much smoother migration, like the one
through a proto-planetary disc, should preserve a good spin-orbit
alignment and, assuming that the axis of the proto-planetary disk
remains constant with time, imply a final hot-Jupiter population
with $\psi \simeq 0$.
%(\textbf{an exhaustive summary of the current theories that
%explain the existence of planets on misaligned orbits can be found
%in Appendix\,\ref{Appendix_A}.}

The experimental measurement of $\psi$ for a large sample of hot
Jupiters is mandatory to guide our comprehension of such planetary
evolution processes in the meanders of theoretical speculations.
Unfortunately, this quantity is very difficult to determine, and
only a few measurements exist (e.g.,
\citealp{brothwell:2014,lendl:2014,lund:2014}). However, its sky
projection, $\lambda$, is a quantity which is commonly measured
for stars hosting \emph{transiting exoplanets} through the
observation of the Rossiter-McLaughlin (RM) effect with
high-precision radial-velocity (RV) instruments. Such measurements
have revealed a wide range of obliquities in both early- and
late-type stars (e.g., \citealp{triaud:2010,albrecht:2012b}),
including planets revolving perpendicular (e.g.,
\citealp{albrecht:2012a}) or retrogade (e.g.,
\citealp{anderson:2010,hebrard:2011,esposito:2014}) to the
direction of the rotation of their parent stars. The results
collected so far are not enough to provide robust statistics and
more investigations are needed to infer a comprehensive picture for
identifying what the most plausible hot-Jupiter migration channels
are.

Within the framework of \texttt{GAPS} (Global Architecture of
Planetary Systems), a manifold long-term observational programme,
using the high-resolution spectrograph HARPS-N at the 3.58\,m TNG
telescope for several semesters
\citep{desidera:2013,desidera:2014,damasso:2015}, we are carrying
out a subprogramme aimed at studying the spin-orbit alignment of a
large sample of known transiting extrasolar planet (TEP) systems
\citep{covino:2013}. This subprogramme is supported by photometric
follow-up observations of planetary-transit events of the targets
in the sample list by using an array of medium-class telescopes.
Monitoring new transits is useful for getting additional
information, as stellar activity, and for constraining the
whole set of physical parameters of the TEP systems better. When
possible, photometric observations are simultaneously performed
with the measurement of the RM effect, benefiting from the
two-telescope observational strategy \citep{ciceri:2013}.

Here, we present the results for two TEP systems: HAT-P-36 and
WASP-11/HAT-P-10.

The paper is organized as follows. In Sect.\,\ref{sec_1.1} we
briefly describe the two systems, subjects of this research study.
The observations and reduction procedures are treated in
Sect.\,\ref{sec_2}, while Sect.\,\ref{sec_3} is dedicated to the
analysis of the photometric data and the refining of the orbital
ephemerides; anomalies detected in the HAT-P-36 light curves are
also discussed. In Sect.\,\ref{sec_4} the HARPS-N time-series data
are used to determine the stellar atmospheric properties and
measure the spin-orbit relative orientation of both the systems.
Sect.\,\ref{sec_5} is devoted to the revision of the main physical
parameters of the two planetary systems, based on the data
previously presented. General empirical correlations between the
orbital obliquity and various properties of TEP systems are
discussed in Sect.\,\ref{sec_6}. The results of this work are
finally summarised in Sect.\,\ref{sec_7}.

% Sect. 1.1
%%%%%%%%%%%%%%%%%%%%%%%%%%%%%%%%%%%%%%%%%%%%%%%%%%%%%%
\subsection{Case history}
\label{sec_1.1}
%%%%%%%%%%%%%%%%%%%%%%%%%%%%%%%%%%%%%%%%%%%%%%%%%%%%%%

HAT-P-36, discovered by the HAT-Net survey \citep{bakos:2012}, is
composed of a $V=12.5$\,mag, G5\,V star (this work), similar to
our Sun, and a hot Jupiter (mass $\sim 1.8 \, M_{\mathrm{Jup}}$
and radius $\sim 1.3 \, R_{\mathrm{Jup}}$), which revolves around
its parent star
%on a slightly eccentric orbit %
in nearly $1.3$\,days. The only follow-up study of this system has
provided a refinement of the transit ephemeris based on the
photometric observation of two planetary transits performed with a
0.6\,m telescope \citep{maciejewski:2013}.

%%%%%%%%%%%%%%%%%%%%%%%%%%%%%%%%%%%%%%%%%%%%%%%%%%%%%%
The discovery of the WASP-11/HAT-P-10 TEP system was almost
concurrently announced by the WASP \citep{west:2009} and HAT-Net
\citep{bakos:2009} teams. It is composed of a $V=11.9$\,mag, K3\,V
star \citep{ehrenreich:2011} (mass $\sim 0.8 \, M_{\sun}$ and
radius $\sim 0.8 \, R_{\sun}$) and a Jovian planet (mass $\sim 0.5
\, M_{\mathrm{Jup}}$ and radius $\sim 1 \, R_{\mathrm{Jup}}$),
moving on a 3.7\,day circular orbit. Four additional transit
observations of this target were obtained by \citet{sada:2012} in
$z^{\prime}$ and $J$ bands with a 0.5\,m and 2.1\,m telescope,
respectively, which were used to refine the transit ephemeris and
photometric parameters. In a more recent study \citep{wang:2014},
an additional four new light curves, which were obtained with a 1\,m
telescope, were presented and most of the physical parameters were
revised, confirming the results of the two discovery papers.

% Sect. 2
%%%%%%%%%%%%%%%%%%%%%%%%%%%%%%%%%%%%%%%%%%%%%%%%%%%%%%
\section{Observation and data reduction} \label{sec_2}
%%%%%%%%%%%%%%%%%%%%%%%%%%%%%%%%%%%%%%%%%%%%%%%%%%%%%%
In this section we present new spectroscopic and photometric
follow-up observations of HAT-P-36 and WASP-11/HAT-P-10. For both
the planetary systems, the RM effect was measured for the first
time and three complete transit light curves were acquired.

% Sect. 2.1
%%%%%%%%%%%%%%%%%%%%%%%%%%%%%%%%%%%%%%%%%%%%%%%%%%%%%%
\subsection{HARPS-N spectroscopic observations}
\label{sec_2.1}
%%%%%%%%%%%%%%%%%%%%%%%%%%%%%%%%%%%%%%%%%%%%%%%%%%%%%%

%%%%%%%%%%%%%%%%%%%%%%%%%%%%%%%%%%%%%%%%%%%%%%%%%%%%%%%
%% Figure: HAT-P-36 RM plot
%%%%%%%%%%%%%%%%%%%%%%%%%%%%%%%%%%%%%%%%%%%%%%%%%%%%%%%
\begin{figure}
\centering
\includegraphics[width=10.5cm]{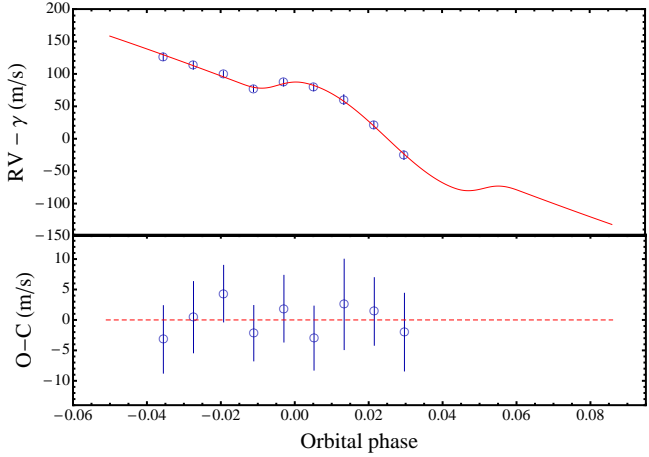}
\caption{Phase-folded RV data of a partial transit of HAT-P-36\,b
observed with HARPS-N. Superimposed is the best-fitting RV-curve
model. The corresponding residuals are plotted in the lower panel.
Phase $0$ corresponds to the time of the planet passing the
periastron.} \label{hatp36RM}
\end{figure}

%%%%%%%%%%%%%%%%%%%%%%%%%%%%%%%%%%%%%%%%%%%%%%%%%%%%%%%
%% Figure: WASP-11 RM plot
%%%%%%%%%%%%%%%%%%%%%%%%%%%%%%%%%%%%%%%%%%%%%%%%%%%%%%%
\begin{figure}
\centering
\includegraphics[width=10.5cm]{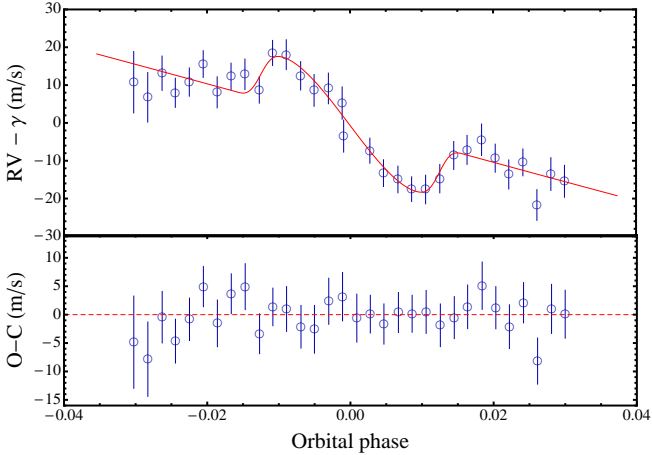}
\caption{Phase-folded RV data of a complete transit of WASP-11\,b
observed with HARPS-N. Superimposed is the best-fitting RV-curve
model. The corresponding
residuals are plotted in the lower panel.} %
\label{wasp11RM}
\end{figure}

Spectroscopic observations of the two targets were performed using
the HARPS-N (High Accuracy Radial velocity Planet Searcher-North;
\citealp{cosentino:2012}) spectrograph at the 3.58\,m TNG, in the
framework of the above-mentioned  \texttt{GAPS} programme.

A sequence of spectra of HAT-P-36 was acquired on the night 2013
February 21 during a transit event of its planet, with an exposure
time of 900\,sec. Unfortunately, because of high humidity, the
observations were stopped before the transit was over and thus
only nine measurements were obtained that cover a bit more than the
first half of the transit.

A complete spectroscopic transit of WASP-11/HAT-P-10\,b was
recorded on 2014 October 2. A sequence of 32 spectra was obtained
using an exposure time of 600\,sec.

The spectra were reduced using the latest
version of the HARPS-N instrument Data Reduction Software
pipeline. The pipeline also provides rebinned 1D spectra that we
used for stellar characterization (see summary
Table\,\ref{tab:hatp36_final_parameters} and
\ref{tab:wasp11_final_parameters}), in particular to estimate the
stellar effective temperature, $T_{\mathrm{eff}}$, and metal
abundance [Fe/H] (see Sect.\,\ref{sec_4.1}). Radial velocities
were measured by applying the weighted cross-correlation function
(CCF) method \citep{baranne:1996,pepe:2002} and using a G2 and a
K5 mask for HAT-P-36 and WASP-11/HAT-P-10, respectively. They are
reported in Table\,\ref{tab:RV} and plotted in
Figs.\,\ref{hatp36RM} and \ref{wasp11RM}, in which we can
immediately note that the RM effect was successfully observed. It
indicates low orbital obliquity for both the systems.

%%%%%%%%%%%%%%%%%%%%%%%%%%%%%%%%%%%%%%%%%%%%%%%%%%%%%
% Table -- HARPS_N RV measurements
%%%%%%%%%%%%%%%%%%%%%%%%%%%%%%%%%%%%%%%%%%%%%%%%%%%%%%
\begin{table}
\caption{HARPS-N RV measurements of HAT-P-36 and WASP-11/HAT-P-10.} %
\label{tab:RV} %
\centering     %
\tiny          %
\setlength{\tabcolsep}{8pt}
\begin{tabular}{lcc}
\hline\hline\\
BJD(TDB) & RV\,(m\,s$^{-1}$)   & Error\,(m\,s$^{-1}$)  \\
%        &         & (m\,s$^{-1}$) & (m\,s$^{-1}$)  \\
\hline\\[-6pt]
\multicolumn{2}{l}{\textbf{HAT-P-36:}}              \\[2pt]%
2\,456\,345.583549   & $-$16200.7  &  5.5   \\
2\,456\,345.594369   & $-$16213.9  &  5.8   \\
2\,456\,345.605189   & $-$16227.2  &  4.6   \\
2\,456\,345.616014   & $-$16249.9  &  4.6   \\
2\,456\,345.626843   & $-$16240.2  &  5.5   \\
2\,456\,345.637659   & $-$16247.8  &  5.3   \\
2\,456\,345.648483   & $-$16266.6  &  7.4   \\
2\,456\,345.659304   & $-$16306.0  &  5.5   \\
2\,456\,345.670130   & $-$16352.7  &  6.4   \\ [6pt] %
%\hline %
\multicolumn{2}{l}{\textbf{WASP-11/HAT-P-10:}} \\[2pt]%
  2\,456\,933.502525 &   4907.4  & 8.0  \\
  2\,456\,933.509760 &   4901.6  & 6.5  \\
  2\,456\,933.516999 &   4907.8  & 4.5  \\
  2\,456\,933.524234 &   4902.7  & 3.8  \\
  2\,456\,933.531469 &   4905.5  & 3.7  \\
  2\,456\,933.538699 &   4910.1  & 3.5  \\
  2\,456\,933.545929 &   4902.3  & 4.1  \\
  2\,456\,933.553165 &   4907.2  & 3.5  \\
  2\,456\,933.560400 &   4908.2  & 4.0  \\
  2\,456\,933.567644 &   4902.8  & 3.5  \\
  2\,456\,933.574884 &   4913.1  & 3.3  \\
  2\,456\,933.582122 &   4913.4  & 3.9  \\
  2\,456\,933.589366 &   4907.7  & 3.7  \\
  2\,456\,933.596598 &   4903.3  & 4.2  \\
  2\,456\,933.603834 &   4904.6  & 4.0  \\
  2\,456\,933.611071 &   4899.7  & 4.2  \\
  2\,456\,933.618316 &   4891.2  & 4.2  \\
  2\,456\,933.625565 &   4887.1  & 3.3  \\
  2\,456\,933.632809 &   4880.9  & 3.5  \\
  2\,456\,933.640053 &   4879.5  & 3.5  \\
  2\,456\,933.647298 &   4877.8  & 3.2  \\
  2\,456\,933.654533 &   4877.9  & 3.8  \\
  2\,456\,933.661763 &   4879.4  & 3.7  \\
  2\,456\,933.669002 &   4885.9  & 3.7  \\
  2\,456\,933.676238 &   4887.0  & 3.8  \\
  2\,456\,933.683469 &   4891.0  & 4.1  \\
  2\,456\,933.690700 &   4884.7  & 3.7  \\
  2\,456\,933.697935 &   4881.6  & 3.8  \\
  2\,456\,933.705166 &   4884.5  & 3.5  \\
  2\,456\,933.712397 &   4873.5  & 4.0  \\
  2\,456\,933.719632 &   4881.2  & 4.3  \\
  2\,456\,933.726867 &   4879.8  & 4.2  \\
\hline
\end{tabular}
%\tablefoot{}
\end{table}
%%%%%%%%%%%%%%%%%%%%%%%%%%%%%%%%%%%%%%%%%%%%%%%%%%%%%%

% Sect. 2.2
%%%%%%%%%%%%%%%%%%%%%%%%%%%%%%%%%%%%%%%%%%%%%%%%%%%%%%%%%%%%
\subsection{Photometric follow-up observations of HAT-P-36}
\label{sec_2.2}
%%%%%%%%%%%%%%%%%%%%%%%%%%%%%%%%%%%%%%%%%%%%%%%%%%%%%%%%%%%%

%%%%%%%%%%%%%%%%%%%%%%%%%%%%%%%%%%%%%%%%%%%%%%%%%%%%%%%
%% Figure: HAT-P-36 Transit light curves
%%%%%%%%%%%%%%%%%%%%%%%%%%%%%%%%%%%%%%%%%%%%%%%%%%%%%%%
\begin{figure*}
\centering
\includegraphics[width=18cm]{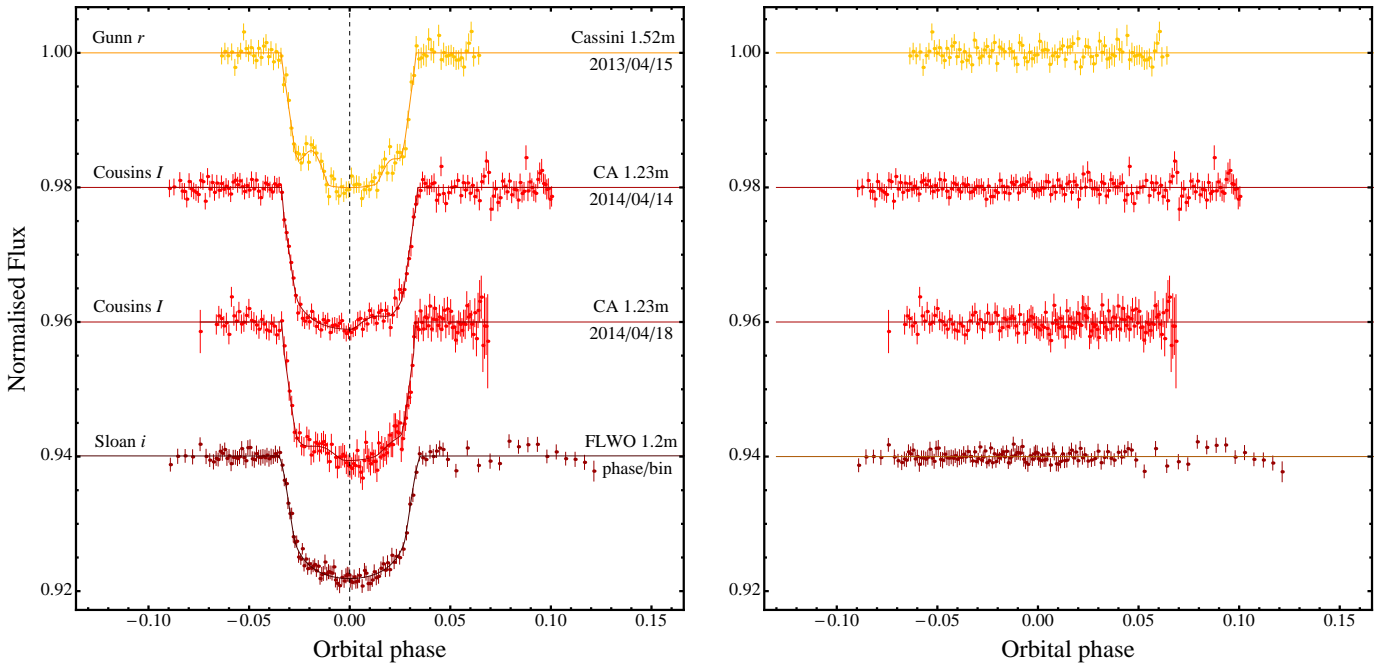}
\caption{Light curves of HAT-P-36\,b transits compared with the
best {\sc prism+gemc} fits. The dates, the telescopes and the
filters used for each transit event are indicated. Residuals from
the fits are displayed in the {\it right panel}. Starting from the top,
the first three light curves (this work) present anomalies that we
interpret as occultations of starspot complexes by the planet.
The third light curve is related to the transit event
successive to that of the second one. The last light curve was
extracted from \citet{bakos:2012} and reported here for
comparison. Here phase 0 corresponds to the mid-time of the
transit.} \label{hatp36lc}
\end{figure*}
%
%%%%%%%%%%%%%%%%%%%%%%%%%%%%%%%%%%%%%%%%%%%%%%%%%%%%%%%
%% Figure: WASP-11 Transit light curves
%%%%%%%%%%%%%%%%%%%%%%%%%%%%%%%%%%%%%%%%%%%%%%%%%%%%%%%
\begin{figure*}
\centering
\includegraphics[width=18cm]{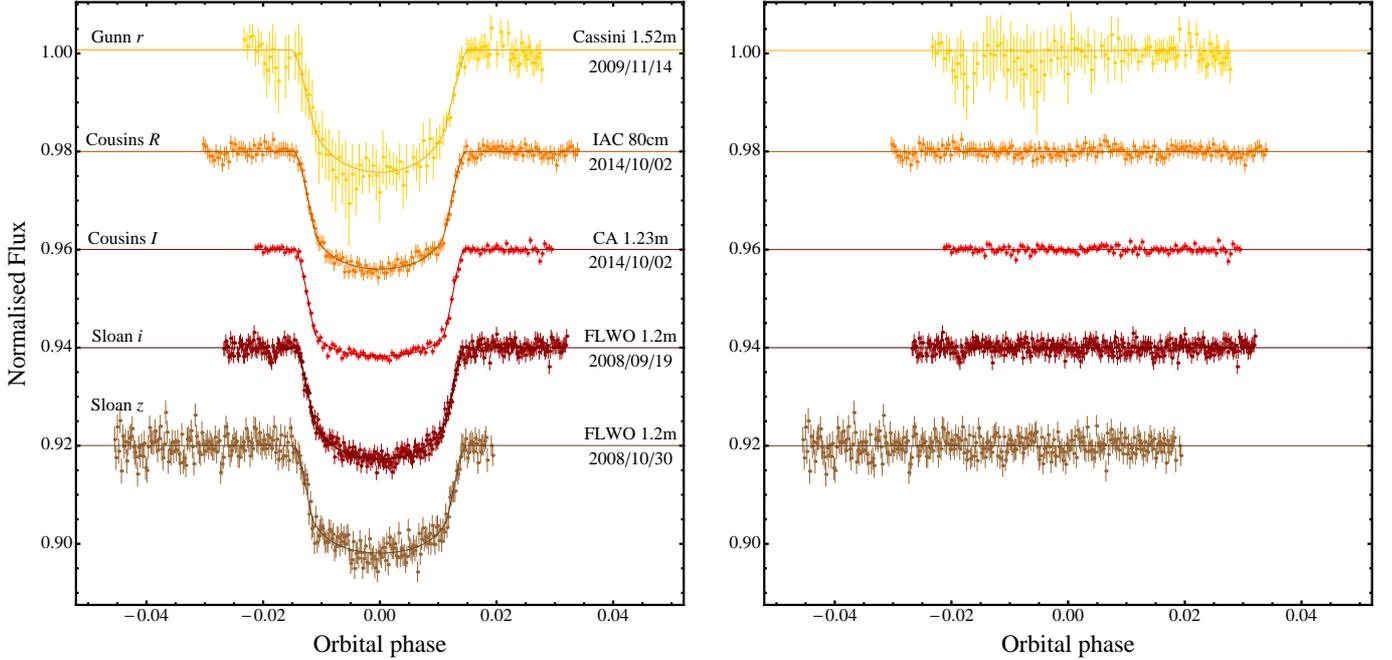}
\caption{Light curves of WASP-11/HAT-P-10\,b transits compared
with the best {\sc jktebop} fits. The dates, the telescopes and
the filters used for each transit event are indicated. Residuals
from the fits are displayed in the {\it right panel}. Starting from the
top, the first three light curves are from this work, whereas the
last two are from \citet{bakos:2009}. The second and
third light curves are related to the same transit event.} %
\label{wasp11lc}
\end{figure*}
%%%%%%%%%%%%%%%%%%%%%%%%%%%%%%%%%%%%%%%%%%%%%%%%%%%%%%

One transit of HAT-P-36\,b was observed on April 2013 through a
Gunn-$r$ filter with the Bologna Faint Object Spectrograph \&
Camera (BFOSC) imager mounted on the 1.52\,m Cassini Telescope at
the Astronomical Observatory of Bologna in Loiano, Italy. The CCD
was used unbinned, giving a plate scale of 0.58 arcsec
pixel$^{-1}$, for a total FOV of 13 arcmin $\times$ 12.6 arcmin.
Two successive transits were observed on April 2014 by using the
Zeiss 1.23\,m telescope at the German-Spanish Astronomical Center
at Calar Alto, Spain. This telescope is equipped with the
DLR-MKIII camera, which has 4000 $\times$ 4000 pixels, a plate
scale of 0.32 arcsec pixel$^{-1}$ and a FOV of 21.5 arcmin
$\times$ 21.5 arcmin. The observations were performed remotely, a
Cousins-$I$ filter was adopted and the CCD was used unbinned, too.
Details of the observations are reported in Table\,\ref{tab:logs}.
As in previous uses of the two telescopes (e.g,
\citealp{mancini:2013a}), the \emph{defocussing} technique was
adopted in all the observations to improve the quality of the
photometric data remarkably. Telescopes were also autoguided.

%%%%%%%%%%%%%%%%%%%%%%%%%%%%%%%%%%%%%%%%%%%%%%%%%%%%%%
% Table -- Night logs
%%%%%%%%%%%%%%%%%%%%%%%%%%%%%%%%%%%%%%%%%%%%%%%%%%%%%%
\begin{table*}
\caption{Details of the transit observations presented in this work.} %
\label{tab:logs} %
\centering     %
\tiny          %
\setlength{\tabcolsep}{5pt}
\begin{tabular}{lccccccccccc}
\hline\hline\\[-6pt]
Telescope & Date of   & Start time & End time  &$N_{\rm obs}$ & $T_{\rm exp}$ & $T_{\rm obs}$ & Filter & Airmass & Moon & Aperture   & Scatter \\
          & first obs &    (UT)    &   (UT)    &              & (s)           & (s)           &        &         &illum.& radii (px) & (mmag)  \\
\hline \\[-6pt]
\multicolumn{10}{l}{\textbf{HAT-P-36:}} \\
Cassini     & 2013 04 14 & 19:58 & 00:39 & 112 & 120 & 144        & Gunn    $r$ & $1.05 \rightarrow 1.000 \rightarrow 1.17$ &  12\% & 16,45,65 & 1.12 \\
CA\,1.23\,m & 2014 04 14 & 19:50 & 02:08 & 176 & 80-120 & 92-132  & Cousins $I$ & $1.28 \rightarrow 1.009 \rightarrow 1.20$ &  98\% & 22,50,75 & 1.10 \\
CA\,1.23\,m & 2014 04 18 & 19:54 & 00:26 & 146 & 80-130 & 92-142 & Cousins $I$ & $1.22 \rightarrow 1.009 \rightarrow 1.06$ &  92\% & 22,50,75 & 1.30 \\ [6pt] %
\multicolumn{10}{l}{\textbf{WASP-11/HAT-P-10:}} \\
Cassini     & 2009 11 14 & 23:02 & 03:36 & 101 & 90-150 & 114-174 & Gunn    $r$ & $1.06 \rightarrow 1.02 \rightarrow 2.00$   &  9\% & 17,25,45 & 2.24 \\
CA\,1.23\,m & 2014 10 02 & 00:51 & 05:23 & 125 & 120    & 132     & Cousins $I$ & $1.12 \rightarrow 1.04 \rightarrow 1.21$   & 52\% & 23,33,60 & 0.65 \\
IAC\,80\,cm & 2014 10 01 & 23:56 & 05:39 & 179 & 90     & 115     & Cousins $R$ & $1.008 \rightarrow 1.000 \rightarrow 2.49$ & 52\% & 28,60,90 & 0.92 \\
\hline
\end{tabular}
\tablefoot{$N_{\rm obs}$ is the number of observations, $T_{\rm
exp}$ is the exposure time, $T_{\rm obs}$ is the observational
cadence, and `Moon illum.' is the geocentric fractional
illumination of the Moon at midnight (UT). The aperture sizes are
the radii of the software apertures for the star, inner sky and
outer sky, respectively. Scatter is the \emph{rms} scatter of the
data versus a fitted model.}
\end{table*}
%%%%%%%%%%%%%%%%%%%%%%%%%%%%%%%%%%%%%%%%%%%%%%%%%%%%%%

The data were reduced with a modified version of the
\texttt{DEFOT} pipeline, written in IDL\footnote{The acronym IDL
stands for Interactive Data Language and is a trademark of ITT
Visual Information Solutions.}, which is exhaustively described in
\citet{southworth:2014}. In brief, we created master calibration
frames by median-combining individual calibration images and used
them to correct the scientific images. A reference image was
selected in each dataset and used to correct pointing variations.
The target and a suitable set of non-variable comparison stars
were identified in the images and three rings were placed
interactively around them; the aperture radii were chosen based on
the lowest scatter achieved when compared with a fitted model.
Differential photometry was finally measured using the
\texttt{APER} routine\footnote{\texttt{APER} is part of the
\texttt{ASTROLIB} subroutine library distributed by NASA.}. Light
curves were created with a second-order polynomial fitted to the
out-of-transit data. The comparison star weights and polynomial
coefficients were fitted simultaneously in order to minimise the
scatter outside transit. Final light curves are plotted in
Fig.\,\ref{hatp36lc} together with one, obtained by
phasing/binning the four light curves reported in the discovery
paper \citep{bakos:2012}, shown here just for comparison.

Very interestingly, the light curves present anomalies
attributable to the passage of the planetary shadow over starspots
or starspot complexes on the photosphere of the host star, which
show up in weak and strong fashion. In particular, the transit
observed with the Cassini telescope was also observed by the
Toru\'{n} 0.6\,m telescope \citep{maciejewski:2013}, which
confirms the presence of anomalies in the light curves.

%%%%%%%%%%%%%%%%%%%%%%%%%%%%%%%%%%%%%%%%%%%%%%%%%%%%%%%%%%%%%%%%%%%

% Sect. 2.3
%%%%%%%%%%%%%%%%%%%%%%%%%%%%%%%%%%%%%%%%%%%%%%%%%%%%%%%%%%%%%%%%%%%
\subsection{Photometric follow-up observations of WASP-11/HAT-P-10}
\label{sec_2.3}
%%%%%%%%%%%%%%%%%%%%%%%%%%%%%%%%%%%%%%%%%%%%%%%%%%%%%%%%%%%%%%%%%%%
A transit observation of WASP-11/HAT-P-10\,b was performed on
November 2009 with the Cassini 1.52\,m telescope through a
Gunn-$r$ filter. Unfortunately, the weather conditions were not
optimal and the data were badly affected by the poor transparency
of the sky. The same transit monitored by HARPS-N (October 2,
2014) was also observed by the Zeiss 1.23\,m telescope through a
Cousins-$I$ filter and a more adequate seeing ($\sim
0.7^{\prime\prime}$). The characteristics of the two telescopes
and observation \emph{modus operandi} were already described in
the previous section. Moreover, the identical transit was observed
with a third telescope, the IAC\,80\,cm telescope, located
at the Teide Observatory on the island of Tenerife (Spain). The
optical channel of this telescope is equipped with CAMELOT, a
$2048 \times 2048$ CCD detector with a plate scale of 0.304
arcsec\,pixel$^{-1}$, providing an astronomical FOV of
10.4\,arcmin $\times$ 10.4\,arcmin. Details of these three
WASP-11/HAT-P-10\,b transit observations are reported in
Table\,\ref{tab:logs}. The data sets were reduced in the same way
as those for the HAT-P-36 case (Sect.\,\ref{sec_2.2}) and the
corresponding light curves are shown in Fig.\,\ref{wasp11lc},
together with other two taken from \citet{bakos:2009}. The
simultaneous transit observation is highlighted in
Fig.\,\ref{wasp11com}, and it allowed an extremely precise measurement
of the mid-point transit time, i.e. $T_{0}=2\,456\,933.615069 \pm
0.000075$\,BJD(TDB).

%%%%%%%%%%%%%%%%%%%%%%%%%%%%%%%%%%%%%%%%%%%%%%%%%%%%%%%%%%%%%%%%%%%

%%%%%%%%%%%%%%%%%%%%%%%%%%%%%%%%%%%%%%%%%%%%%%%%%%%%%%%
%% Figure: WASP-11 combined light curve
%%%%%%%%%%%%%%%%%%%%%%%%%%%%%%%%%%%%%%%%%%%%%%%%%%%%%%%
\begin{figure*}
\centering
\includegraphics[width=18.cm]{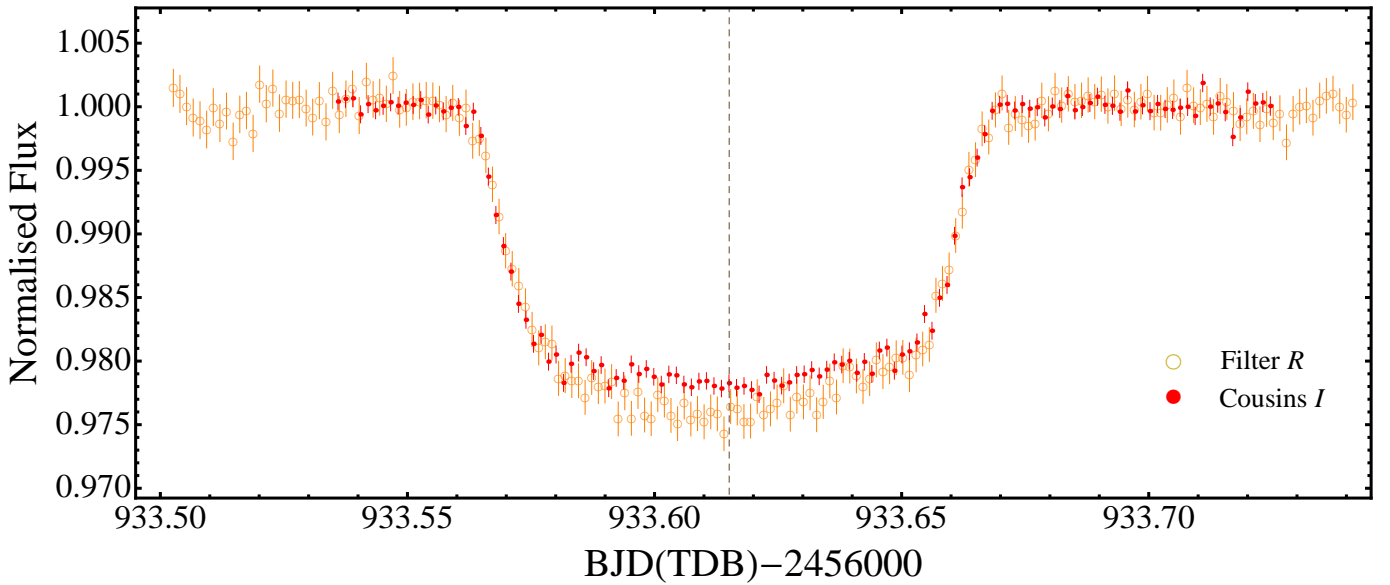}
\caption{Light curve of a transit event of WASP-11/HAT-P-10\,b
observed simultaneously with the CA 1.23\,m (red points) and the
IAC 80\,cm (empty orange circles) telescopes. The data of the two
telescopes are superimposed to highlight the difference in the
light-curve shape caused by the different filter adopted
\citep{knutson:2007}. The dashed vertical line represents the time
of transit mid-point that, thanks to the two-site strategy, was
very accurately measured.
%Both the data sets show a slight
%asymmetry (compare the first with the second part of the transit),
%probably caused by stellar activity.
}
\label{wasp11com} %
\end{figure*}

% Sect. 3
%%%%%%%%%%%%%%%%%%%%%%%%%%%%%%%%%%%%%%%%%%%%%%%%%%%%%%%%%%%%%%%%%%%
\section{Light-curve analysis}
\label{sec_3}
%%%%%%%%%%%%%%%%%%%%%%%%%%%%%%%%%%%%%%%%%%%%%%%%%%%%%%%%%%%%%%%%%%%
The three HAT-P-36 light curves show possible starspot crossing
events, which must be analysed using a self-consistent and
physically realistic model. As in previous cases
\citep{mancini:2013b,mancini:2014b,mohler:2013}, we utilise the
{\sc prism}\footnote{Planetary Retrospective Integrated Star-spot
Model.} and {\sc gemc}\footnote{Genetic Evolution Markov Chain.}
codes \citep{tregloan:2013,tregloan:2015} to undertake this task.
{\sc prism} performs a modelling of planetary-transit light curves
with one or more starspots by means of a pixellation approach in
Cartesian coordinates. {\sc gemc} uses a Differential Evolution
Markov Chain Monte Carlo (DE-MCMC) approach to locate the
parameters of the {\sc prism} model that better fit the data,
using a global search. The fitted parameters of {\sc prism} are
the sum and ratio of the fractional radii\footnote{The fractional
radii are defined as $r_{\mathrm{A}} = \frac{R_{\star}}{a}$ and
$r_{\mathrm{b}} = \frac{R_{\mathrm{p}}}{a}$, where $R_{\star}$ and
$R_{\mathrm{p}}$ are the true radii of the star and planet, and
$a$ is the orbital semimajor axis.}
($r_{\mathrm{A}}+r_{\mathrm{b}}$ and
$k=\frac{r_{\mathrm{b}}}{r_{\mathrm{A}}}$), the orbital period and
inclination ($P$ and $i$), the time of transit midpoint ($T_{0}$)
and the coefficients of the quadratic limb darkening law
($u_{\mathrm{A}}$ and $v_{\mathrm{A}}$).

The {\sc prism}+{\sc gemc} code works in the way that the user has to decide how
many starspots have to be fitted. We fitted for all the spot anomalies that can be seen
in the transit light curves. Each starspot is then
reproduced by the longitude and co-latitude of its centre
($\theta$ and $\phi$), its angular radius ($r_{\mathrm{spot}}$)
and its contrast ($\rho_{\mathrm{spot}}$), the last being the
ratio of the surface brightness of the starspot to that of the
surrounding photosphere. The orbital eccentricity was fixed to
zero \citep{bakos:2012}.

%For planets with eccentric orbits, it is possible to introduce the
%eccentricity, $e$, and the argument of the periastron, $\omega$,
%to better estimate the uncertainties in the system parameters,
%especially for $i$ and $r_{\mathrm{A}}$. {\sc gemc} searches the
%best values in the physical range from $0$ to $1$ for $e$ and from
%$0$ to $360^{\circ}$ for $\omega$, but uses a Gaussian prior to
%constrain the values. For the case of HAT-P-36, we have used the
%values from \citet{bakos:2012}, i.e. $e=0.063 \pm 0.032$
%and $\omega=95.0 \pm 63.0$.

The derived parameters of the planetary system are reported in
Table\,\ref{tab:hatp36_phot_parameters_1}, while those of the
starspots in Table\,\ref{tab:hatp36_phot_parameters_2}. The light
curves and their best-fitting models are shown in
Fig.\,\ref{hatp36lc}. For each transit, a representation of the
starspots on the stellar disc is drawn in
Fig.\,\ref{hatp36starspots}.

The light curves for the WASP-11/HAT-P-10 transits were
modelled by {\sc jktebop}\footnote{\textsc{jktebop} is written in
FORTRAN77 and is available at: {\tt
http://www.astro.keele.ac.uk/jkt/codes/jktebop.html}} (see
\citealp{southworth:2008,southworth:2013} and references therein),
as this code is much faster than the previous one and no clear
starspot anomalies are visible. The parameters used for {\sc
jktebop} were the same as for {\sc prism} and the orbital
eccentricity was also fixed to zero \citep{bakos:2009}. Light
curves from the second discovery paper \citep{bakos:2009} were
also reanalysed. The results of the fits are summarised in
Table\,\ref{tab:wasp11_phot_parameters} and displayed in
Fig.\,\ref{wasp11lc}. Its low quality meant that the Cassini light
curve (top curve in Fig.\,\ref{wasp11lc}) was excluded from the
analysis.

%%%%%%%%%%%%%%%%%%%%%%%%%%%%%%%%%%%%%%%%%%%%%%%%%%%%%%
% Table -- HAT-P-36 gemc results part 1
%%%%%%%%%%%%%%%%%%%%%%%%%%%%%%%%%%%%%%%%%%%%%%%%%%%%%%
\begin{table*}
%\tiny %
\caption{Photometric properties of the HAT-P-36 system derived by
fitting the light curves with {\sc gemc}.}
\label{tab:hatp36_phot_parameters_1} %
\centering %
\begin{tabular}{llccccc}
\hline %
\hline \\[-8pt]%
Source & Filter & $r_{\mathrm{A}}+r_{\mathrm{b}}$& $k$ & $i^{\circ}$ & $u_{\mathrm{A}}$ &$v_{\mathrm{A}}$\\
\hline\\[-6pt]
Cassini\,1.52\,m & Gunn $r$    & $0.2278 \pm 0.0037$ & $0.1243 \pm 0.0025$ & $85.95 \pm 0.34$ & $0.24 \pm 0.13$ & $0.40 \pm 0.21$ \\
CA\,1.23\,m \#1  & Cousins $I$ & $0.2298 \pm 0.0039$ & $0.1290 \pm 0.0023$ & $85.76 \pm 0.36$ & $0.25 \pm 0.10$ & $0.29 \pm 0.19$ \\
CA\,1.23\,m \#2  & Cousins $I$ & $0.2285 \pm 0.0039$ & $0.1298 \pm 0.0012$ & $85.86 \pm 0.43$ & $0.32 \pm 0.11$ & $0.15 \pm 0.14$ \\
\hline
%{\bf Final results}  & $\mathbf{0.24539 \pm 0.00499}$ & $\mathbf{0.11616 \pm 0.00081}$ & $\mathbf{85.74 \pm 0.95}$ & $\mathbf{0.21998 \pm 0.00436}$ & $\mathbf{0.02541 \pm 0.00065}$ \\
%\hline
%\citet{bakos:2012} &    & ...                        & $0.1186 \pm 0.0012$ & $86.0 \pm 0.3$   & $0.063 \pm 0.032$   & $95 \pm 63$      \\
\hline
\end{tabular}
\end{table*}

%%%%%%%%%%%%%%%%%%%%%%%%%%%%%%%%%%%%%%%%%%%%%%%%%%%%%%
% Table -- HAT-P-36 gemc results part 2
%%%%%%%%%%%%%%%%%%%%%%%%%%%%%%%%%%%%%%%%%%%%%%%%%%%%%%
\begin{table*}
\caption{Starspot parameters derived from the {\sc gemc} fitting
of the HAT-P-36 transit light curves presented in this work.}
\label{tab:hatp36_phot_parameters_2} %
\centering %
\begin{tabular}{lcccccc}
\hline
\hline \\[-8pt]%
Telescope & starspot & $\theta (^{\circ})\,^{a}$ & $\phi(^{\circ})\,^{b}$ & $r_{\rm spot}(^{\circ})\,^{c}$ & $\rho_{\rm spot}\,^{d}$ & Temperature (K)$\,^{e}$   \\
\hline  \\[-6pt]%%
Cassini\,1.52\,m & $\begin{tabular}{c}
\#1  \\
\#2  \\
\end{tabular}$ & $\begin{tabular}{c}
$-36.59 \pm 1.85$ \\
$\;\;\,38.29 \pm 2.88$ \\
\end{tabular}$ & $\begin{tabular}{c}
$77.41 \pm 5.82$ \\
$82.68 \pm 5.27$ \\
\end{tabular}$ & $\begin{tabular}{c}
$11.42 \pm 3.45$ \\
$12.54 \pm 5.57$ \\
\end{tabular}$ & $\begin{tabular}{c}
$0.65 \pm 0.07$  \\
$0.62 \pm 0.23$ \\
\end{tabular}$ &
$\begin{tabular}{c}
$5070 \pm 129$  \\
$5016 \pm 421$ \\
\end{tabular}$  \\
\hline  \\[-8pt]%
CA\,1.23\,m \#1  & \#1 & $\;\;\,21.06 \pm 3.20$ & $86.52 \pm 3.96$ & $17.67 \pm 4.81$ & $0.86_{-0.17}^{+0.14}$ & $5380 \pm 305$ \\
\hline  \\[-8pt]%%
CA\,1.23\,m \#2  & $\begin{tabular}{c}
\#1  \\
\#2  \\
\end{tabular}$ & $\begin{tabular}{c}
$-24.99 \pm 4.91$ \\
$\;\;\,46.00 \pm 14.13$ \\
\end{tabular}$ & $\begin{tabular}{c}
$87.08 \pm 3.17$ \\
$88.27 \pm 2.38$ \\
\end{tabular}$ & $\begin{tabular}{c}
$21.38 \pm 6.82$ \\
$18.66 \pm 3.52$ \\
\end{tabular}$ & $\begin{tabular}{c}
$0.92_{-0.14}^{+0.08}$  \\
$0.78 \pm 0.19$ \\
\end{tabular}$ & $\begin{tabular}{c}
$5485 \pm 244$  \\
$5253 \pm 353$ \\
\end{tabular}$  \\
\hline %
\end{tabular}
\tablefoot{$^{(a)}$The longitude of the centre of the spot is
defined to be $0^{\circ}$ at the centre of the stellar disc and
can vary from $-90^{\circ}$ to $90^{\circ}$. $^{(b)}$The
co-latitude of the centre of the spot is defined to be $0^{\circ}$
at the north pole and $180^{\circ}$ at the south pole.
$^{(c)}$Angular radius of the starspot (note that an angular
radius of $90^{\circ}$ covers half of stellar surface).
$^{(d)}$Spot contrast; note that 1.0 equals the brightness of the
surrounding photosphere. $^{(e)}$The temperature of the starspots
are obtained by considering the photosphere and the starspots as
black bodies.}
\end{table*}

%%%%%%%%%%%%%%%%%%%%%%%%%%%%%%%%%%%%%%%%%%%%%%%%%%%%%%
% Table -- WASP-11 jktbop results
%%%%%%%%%%%%%%%%%%%%%%%%%%%%%%%%%%%%%%%%%%%%%%%%%%%%%%
\begin{table*}
\caption{Photometric properties of the WASP-11/HAT-P-10 system
derived by fitting the light curves with {\sc jktebop}.}
\label{tab:wasp11_phot_parameters} %
\centering %
\begin{tabular}{llccccc}
\hline \hline \\[-8pt]%
Source & Filter & $r_{\mathrm{A}}+r_{\mathrm{b}}$& $k$ & $i^{\circ}$ & $u_{\mathrm{A}}$ &$v_{\mathrm{A}}$\\
\hline \\[-8pt]%
Cassini\,1.52\,m & Gunn $r$& $0.10137 \pm 0.01380$ & $0.13639 \pm 0.00868$ & $88.00 \pm 1.97$ & $0.56 \pm 0.23$ & $0.21 \pm 0.06$ \\
CA\,1.23\,m  & Cousins $I$ & $0.09230 \pm 0.00197$ & $0.13234 \pm 0.00078$ & $89.09 \pm 0.68$ & $0.24 \pm 0.04$ & $0.25 \pm 0.07$ \\
IAC\,80\,cm  & Cousins $R$ & $0.09695 \pm 0.00358$ & $0.13657 \pm 0.00200$ & $88.29 \pm 0.54$ & $0.40 \pm 0.08$ & $0.21 \pm 0.07$ \\
FLWO\,1.2\,m & Sloan $z$   & $0.09277 \pm 0.00350$ & $0.12818 \pm 0.00181$ & $89.95 \pm 0.92$ & $0.41 \pm 0.06$ & $0.27 \pm 0.07$ \\
FLWO\,1.2\,m & Sloan $i$   & $0.09160 \pm 0.00242$ & $0.12856 \pm 0.00156$ & $89.94 \pm 0.76$ & $0.49 \pm 0.05$ & $0.25 \pm 0.07$ \\
%{\bf Final results}  & $\mathbf{0.24539 \pm 0.00499}$ & $\mathbf{0.11616 \pm 0.00081}$ & $\mathbf{85.74 \pm 0.95}$ & $\mathbf{0.21998 \pm 0.00436}$ & $\mathbf{0.02541 \pm 0.00065}$ \\
%\hline
%\citet{west:2009}  & & & $0.1273_{-0.0031}^{+0.0047}$    & $89.8_{-0.8}^{+0.2}$    & & \\
%\citet{bakos:2009} & & & $0.1315 \pm 0.0010$             & $88.6_{-0.4}^{+0.5}$    & & \\
%\citet{wang:2014}  & & & $0.13103_{-0.00032}^{+0.00044}$ & $89.14_{-0.47}^{+0.50}$ & & \\
\hline
\end{tabular}
\end{table*}

% Sect. 3.1
%%%%%%%%%%%%%%%%%%%%%%%%%%%%%%%%%%%%%%%%%%%%%%%%%%%%%%%%%%%%%%%%%%%
\subsection{Orbital period determination} %
\label{sec_3.1}
%%%%%%%%%%%%%%%%%%%%%%%%%%%%%%%%%%%%%%%%%%%%%%%%%%%%%%%%%%%%%%%%%%%

%%%%%%%%%%%%%%%%%%%%%%%%%%%%%%%%%%%%%%%%
%% Table HAT-P-36 O-C plot
%%%%%%%%%%%%%%%%%%%%%%%%%%%%%%%%%%%%%%%%
\begin{table}
\caption{Times of transit midpoint of HAT-P-36\,b and their
residuals.}
\label{tab:hatp36oc} %
\centering %
\begin{tabular}{lrrc}
\hline
\hline  \\[-8pt]%%
Time of minimum    & Cycle & O-C & Reference  \\
BJD(TDB)$-2400000$ & no.   & (JD)     &            \\
\hline \\[-8pt]%
$55555.89060    \pm  0.00043$ &     -7   &  0.000376 &  1   \\
$55597.04088    \pm  0.00770$ &     24   &  0.002904 &  1   \\
$55601.01882    \pm  0.00074$ &     27   & -0.001197 &  1   \\
$55608.98390    \pm  0.00030$ &     33   & -0.000198 &  1   \\
$56365.56800    \pm  0.00234$ &    603   & -0.001928 &  3   \\
$56397.42170    \pm  0.00202$ &    627   & -0.004655 &  3   \\
$56397.42894    \pm  0.00031$ &    627   &  0.000822 &  2   \\
$56762.44834    \pm  0.00018$ &    902   & -0.000158 &  3   \\
$56766.43055    \pm  0.00028$ &    905   & -0.000011 &  3   \\
\hline
\end{tabular}
\tablefoot{References: (1) FLWO 1.2\,m \citep{bakos:2012}; (2)
Cassini 1.52\,m (this work); (3) CA 1.23\,m (this work).}
\end{table}

We used the new photometric data to refine the transit ephemeris
of HAT-P-36\,b. The transit times and uncertainties were obtained
using {\sc prism+gemc} as previously explained. To these timings,
we added four from the discovery paper \citep{bakos:2012} and two
from \citet{maciejewski:2013}\footnote{We added 1/25 days to the
published values to correct an error caused by a misunderstanding
in reading the time stamps of the fits files.}. The nine timings were
placed on the BJD(TDB) time system (Table\,\ref{tab:hatp36oc}).
The resulting measurements of transit midpoints were fitted with a
straight line to obtain a final orbital ephemeris
%
%\begin{linenomath}
\begin{equation}
 P= 1.32734684 \pm 0.00000050~\rm{d} \nonumber
\end{equation}
and the mid-transit time at cycle zero
\begin{equation}
T_{0}=2455565.18165 \pm 0.00037 ~\rm{BJD(TDB)}\, , \nonumber \;
\end{equation}
%\end{linenomath}
%
with reduced $\chi_{\nu}^2 = 2.5$, which is significantly greater
than unity.
A plot of the residuals around the fit (see Fig.\,\ref{hatp36oc})
does not indicate any clear systematic deviation from the
predicted transit times.

We introduced other 21 timings in the analysis measured based on
transit light curves observed by amateur astronomers and available
on the ETD\footnote{The Exoplanet Transit Database (ETD) website
can be found at http://var2.astro.cz/ETD} website. These 21 light
curves were selected considering if they had complete coverage of
the transit and a Data Quality index $\leq 3$. Repeating the
analysis with a larger sample, we obtained $4.1$ as the reduced
$\chi^2_{\nu}$ of the fit. A high value of $\chi_{\nu}^2$ means that
the assumption of a constant orbital period does not agree with the 
timing measurements. This could indicate either transit
timing variations (TTVs) or that the measurement errors are underestimated.
The latter possibility can occur very easily, but be difficult to rule out, so the
detection of TTVs requires more than just an excess $\chi_{\nu}^2$.
The Lomb-Scargle periodogram generated for timing residuals shows no
significant signal, making any periodic variation unlikely, so we do not claim the
existence of TTVs in this system. The residuals of the ETD timings are also
shown in Fig.\,\ref{hatp36oc} for completeness.

%%%%%%%%%%%%%%%%%%%%%%%%%%%%%%%%%%%%%%%%%%%%%%%%%%%%%%%
%% Figure: HAT-P-36 O-C plot
%%%%%%%%%%%%%%%%%%%%%%%%%%%%%%%%%%%%%%%%%%%%%%%%%%%%%%%
\begin{figure*}
\centering
\includegraphics[width=18cm]{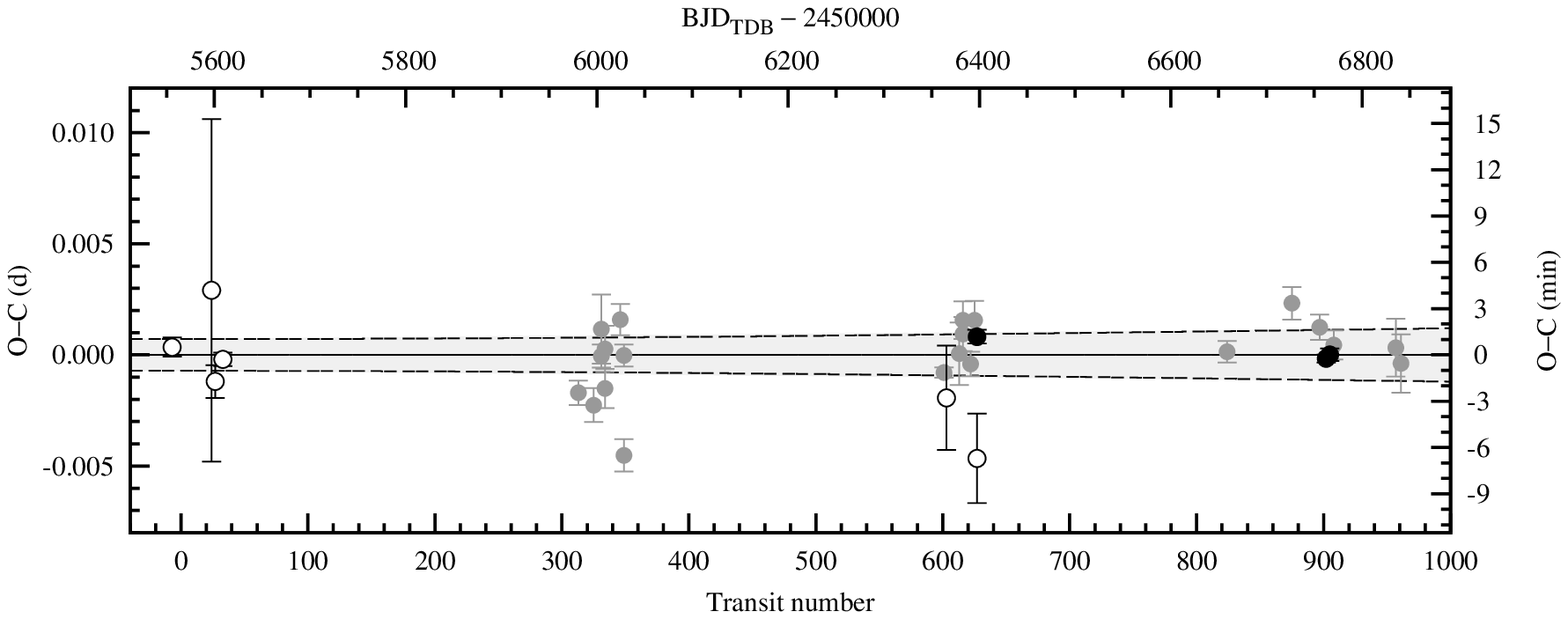}
\caption{Residuals for the timings of HAT-P-36\,b at mid-transit
versus a linear ephemeris. The four timings reported by
\citet{bakos:2012} and the two by \citet{maciejewski:2013} are
plotted using empty circles, while those marked with gray circles
are from ETD. The timings measured by this work are shown with
black circles. The dashed lines mark 2-$\sigma$ uncertainties of
the ephemeris (i.e. at the $95\%$
confidence level), which was estimated without considering the data from ETD.} %
\label{hatp36oc}
\end{figure*}
%%%%%%%%%%%%%%%%%%%%%%%%%%%%%%%%%%%%%%%%%%%%%%%%%%%%%%

%%%%%%%%%%%%%%%%%%%%%%%%%%%%%%%%%%%%%%%%
%% Table WASP-11 O-C plot
%%%%%%%%%%%%%%%%%%%%%%%%%%%%%%%%%%%%%%%%
\begin{table}
\caption{Times of transit midpoint of WASP-11/HAT-P-10\,b and
their residuals.}
\label{tab:wasp11oc} %
\centering %
\begin{tabular}{lrrc}
\hline
\hline  \\[-8pt]%%
Time of minimum    & Cycle & O-C & Reference  \\
BJD(TDB)$-2400000$ & no.   & (JD)     &            \\
\hline \\[-8pt]%
$54729.90657 \pm 0.00023$ & 0   & $-0.000589$ & 1 \\
$54770.85432 \pm 0.00011$ & 11  & $-0.000115$ & 1 \\
$54793.19692 \pm 0.00076$ & 17  & $ 0.002071$ & 2 \\
$55143.10895 \pm 0.00046$ & 111 & $ 0.001013$ & 2 \\
$55161.71529 \pm 0.00021$ & 116 & $ 0.000490$ & 3 \\
$55507.90419 \pm 0.00042$ & 209 & $-0.001219$ & 3 \\
$55842.92921 \pm 0.00021$ & 299 & $ 0.000631$ & 3 \\
$55842.92952 \pm 0.00044$ & 299 & $ 0.000941$ & 3 \\
$55865.26734 \pm 0.00166$ & 305 & $-0.001483$ & 2 \\
$55150.54662 \pm 0.00104$ & 113 & $-0.000741$ & 4 \\
$56933.61504 \pm 0.00008$ & 592 & $-0.000086$ & 5 \\
$56933.61522 \pm 0.00018$ & 592 & $ 0.000099$ & 6 \\
\hline
\end{tabular}
\tablefoot{References: (1) FLWO 1.2\,m \citep{bakos:2009}; (2)
\citet{wang:2014}; (3) \citet{sada:2012}; (4) Cassini 1.52\,m
(this work); (5) CA 1.23\,m (this work); (6) IAC 80\,cm (this
work).}
\end{table}

New mid-transit times (three transits; this work) and those
available in the literature (9 transits) allowed us to refine
transit ephemeris for WASP-11/HAT-P-10\,b too. In particular, the
two light curves from \citet{bakos:2009} were re-fitted with {\sc
jktebop} (Fig.\,\ref{wasp11lc}). All the timings are summarised in
Table\,\ref{tab:wasp11oc}. As a result of a linear fit, in which
timing uncertainties were taken as weights, we derived
%
%\begin{linenomath}
\begin{equation}
 P=3.72247967 \pm 0.00000045~\rm{d} \nonumber
\end{equation}
and the mid-transit time at cycle zero
\begin{equation}
T_{0}=2454729.90716 \pm 0.00020 ~\rm{BJD_{TDB}}\, , \nonumber \;
\end{equation}
%\end{linenomath}
%
with $\chi^2_{\nu}=5.0$, which is again far from unity. This could
be explained again by underestimated timing uncertainties or
hypothesizing a variation in transit times caused by unseen
planetary companion or stellar activity. However, also for this
case the Lomb-Scargle periodogram shows no significant signal, not
supporting the second hypothesis. The plot of the residuals around
the fit is shown in Fig.\,\ref{wasp11oc}.

As in the previous case, we selected 34 transits from ETD, using
the same criteria explained above. Most of the corresponding
timings are very scattered around the predicted transit mid-times,
but have tiny error bars. Including them in the fit returns a
higher value for the $\chi_{\nu}^2$, suggesting that their
uncertainties are very underestimated, and the data reported on ETD
should be used cautiously. They are shown in Fig.\,\ref{wasp11oc}.

%%%%%%%%%%%%%%%%%%%%%%%%%%%%%%%%%%%%%%%%%%%%%%%%%%%%%%%
%% Figure: WASP-11 O-C plot
%%%%%%%%%%%%%%%%%%%%%%%%%%%%%%%%%%%%%%%%%%%%%%%%%%%%%%%
\begin{figure*}
\centering
\includegraphics[width=18cm]{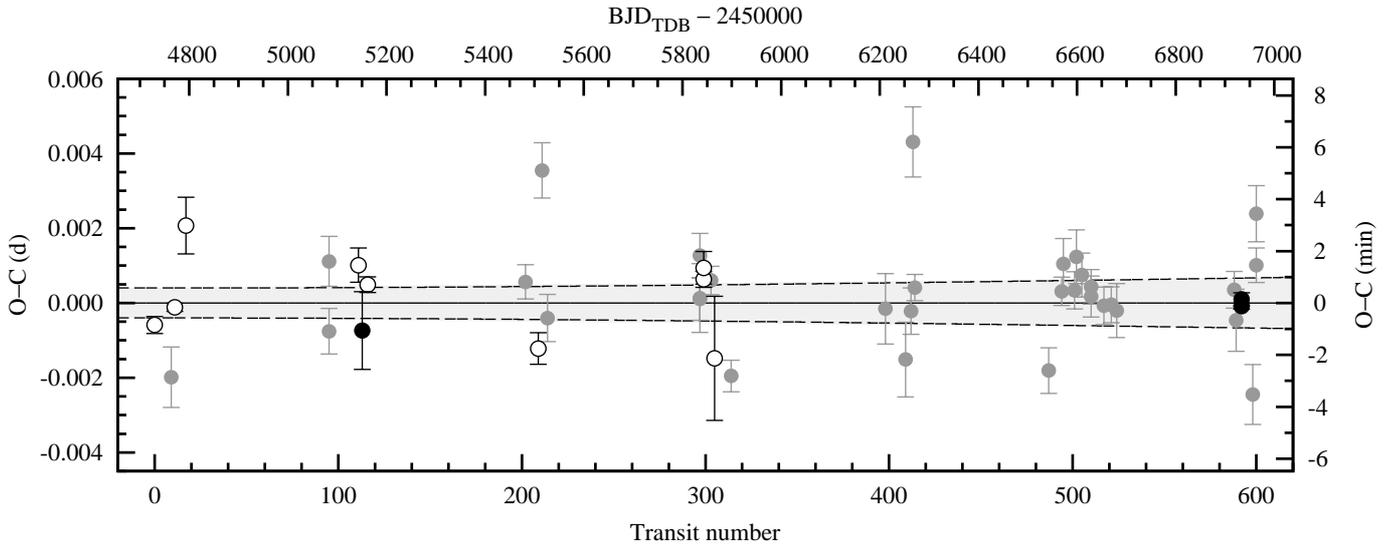}
\caption{Residuals for the timings of WASP-11/HAT-P-10\,b at
mid-transit versus a linear ephemeris. The filled black dots are
from this work, open circles from the literature, (i.e.
\citealp{bakos:2009,sada:2012,wang:2014}) and gray dots from ETD.
The dashed lines mark 2-$\sigma$ uncertainties of the
ephemeris, which was estimated without considering the data from ETD.}%
\label{wasp11oc}
\end{figure*}
%%%%%%%%%%%%%%%%%%%%%%%%%%%%%%%%%%%%%%%%%%%%%%%%%%%%%%

We would like to stress however that, since in both the cases the
timing measurements are in groups separated by several hundred of
days, we are not sensitive to all the periodicities. Only by
performing systematic observations of many subsequent transits it
is possible to rule out the presence of additional bodies in the
two TEP systems with higher confidence.

% Sect. 3.2
%%%%%%%%%%%%%%%%%%%%%%%%%%%%%%%%%%%%%%%%%%%%%%%%%%%%%%%%%%%%%%%%%%%
\subsection{HAT-P-36 starspots} %
\label{sec_3.2}
%%%%%%%%%%%%%%%%%%%%%%%%%%%%%%%%%%%%%%%%%%%%%%%%%%%%%%%%%%%%%%%%%%%

%%%%%%%%%%%%%%%%%%%%%%%%%%%%%%%%%%%%%%%%%%%%%%%%%%%%%%%
%% Figure: HAT-P-36 starspots
%%%%%%%%%%%%%%%%%%%%%%%%%%%%%%%%%%%%%%%%%%%%%%%%%%%%%%%
\begin{figure*}
\centering
\includegraphics[width=18cm]{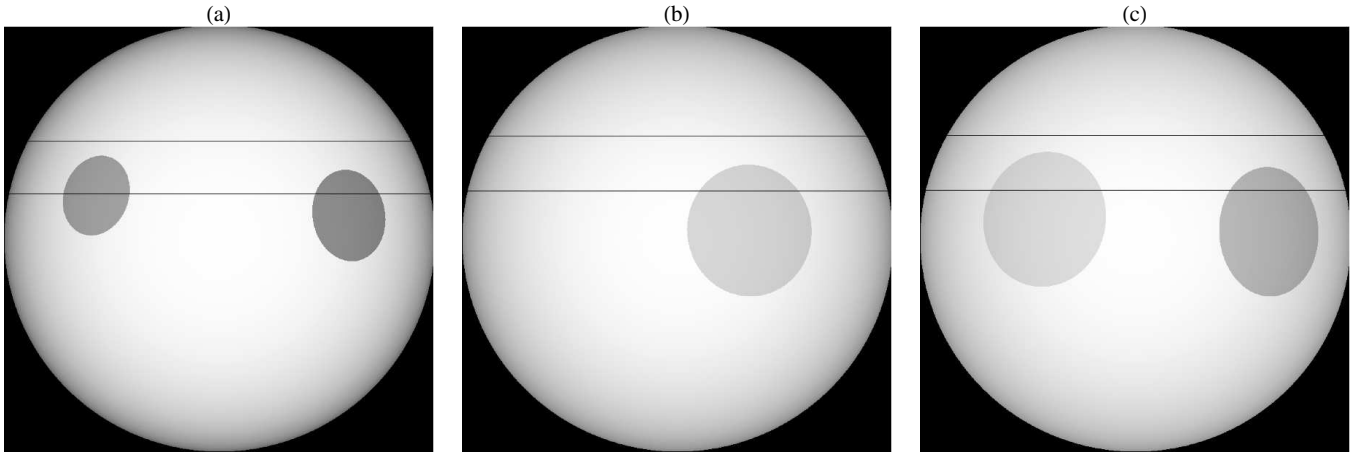}
\caption{Representation of the stellar disc, starspot positions
and transit chord for the three HAT-P-36 datasets containing spot
anomalies: (a) refers to the transit observed with the Cassini
1.52\,m telescope on 2013 April 15, while (b) and (c) to those
observed with the CA 1.23\,m telescope on 2014 April 14 and 18,
respectively. The gray scale of each starspot is related to its
contrast (compare with values in
Table\,\ref{tab:hatp36_phot_parameters_2}).}
\label{hatp36starspots}
\end{figure*}

As described in Sect.\,\ref{sec_3}, the anomalies in the three
HAT-P-36 light curves were modelled as starspots, whose parameters
were fitted together with those of the transits (see
Tables \ref{tab:hatp36_phot_parameters_1} and
\ref{tab:hatp36_phot_parameters_2}). The stellar disc, the
positions of starspots and the transit chords are displayed in
Fig.\,\ref{hatp36starspots}, based on the results of the
modelling. Considering both the photosphere and the starspots as
black bodies \citep{rabus:2009,sanchis:2011,mohler:2013} and using
Eq.\,1 of \citet{silva:2003} and $T_{\mathrm{eff}}=5620\pm40$ (see
Sect.\,\ref{sec_4.1}), we estimated the temperature of the
starspots at different bands and reported them in the last column
of Table \ref{tab:hatp36_phot_parameters_2}. The values of the
starspot temperature estimated in the transit on April 2013 are in
good agreement with each other within the experimental
uncertainties. The same is true for those observed in the two
transits on April 2014, even if they point to starspots with
temperature higher than, but still compatible, with those of the
previous year. The starspot temperatures measured from the Loiano
and CAHA observations are consistent with what has been measured
for other main-sequence stars in transit observations, as we can
see from Fig.\,\ref{starspots_tem_plot}, where we
report the starspot temperature contrast versus the temperature
of the photosphere of the corresponding star for data taken from
the literature. The spectral class of the stars is also reported
and allows us to figure out that the temperature difference
between photosphere and starspots is not strongly dependent on
spectral type, as already noted by \citet{strassmeier2009}.

%%%%%%%%%%%%%%%%%%%%%%%%%%%%%%%%%%%%%%%%%%%%%%%%%%%%%%%
%% Figure: Starspots plot
%%%%%%%%%%%%%%%%%%%%%%%%%%%%%%%%%%%%%%%%%%%%%%%%%%%%%%%
\begin{figure*}
\centering
\includegraphics[width=18cm]{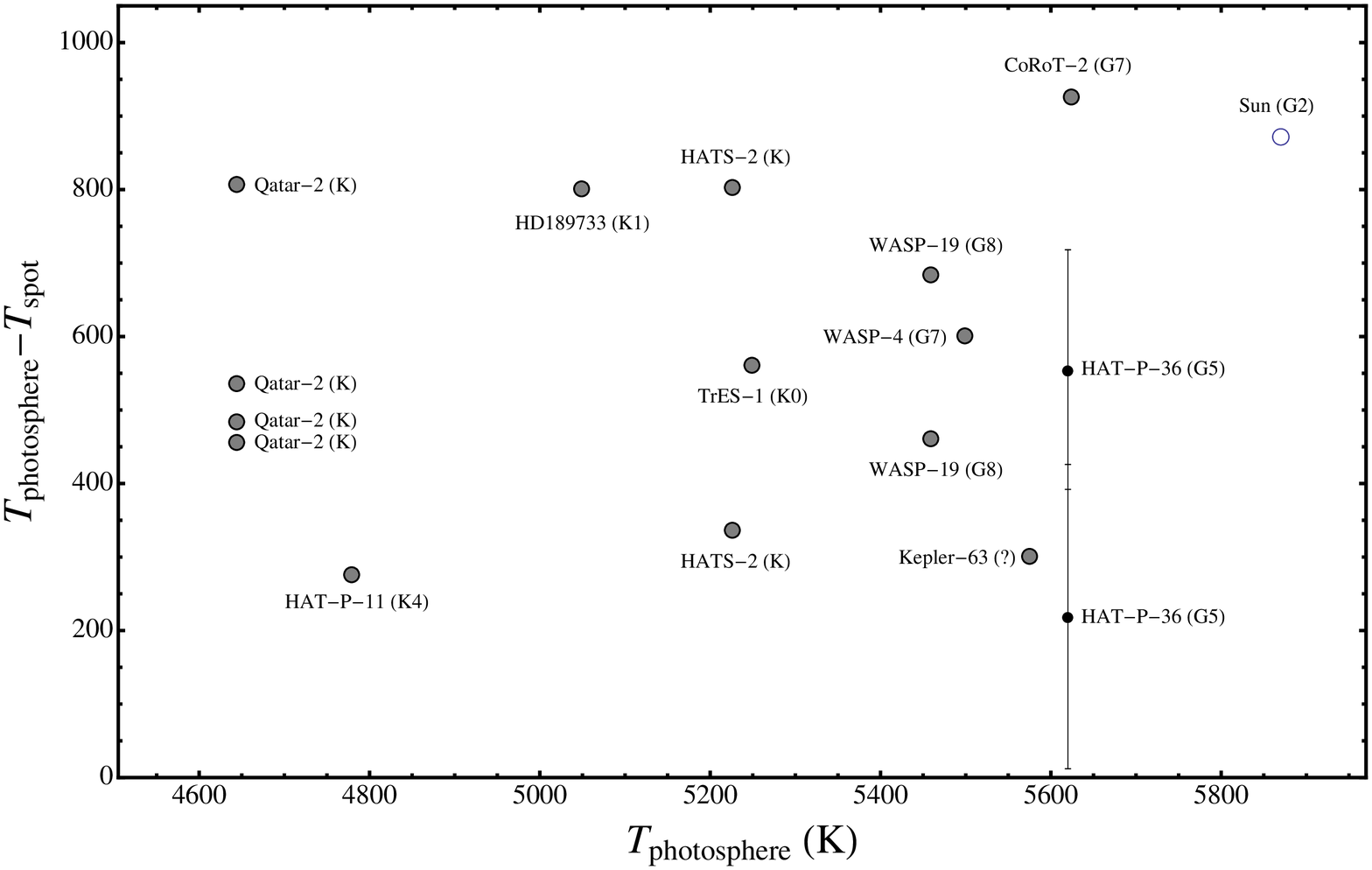}
\caption{Starspot temperature contrast with respect to the
photospheric temperature in several dwarf stars. The name and
spectral type of the star are also reported. The values for
TrES-1, CoRoT-2, HD\,189733, WASP-4, HATS-2, Kepler-63, Qatar-2
are taken from \citet{rabus:2009}, \citet{silva:2010},
\citet{sing:2011}, \citet{sanchis:2011}, \citet{mohler:2013},
\citet{sanchis:2013}, \citet{mancini:2014b}, respectively. The two
values for WASP-19 are from \citet{mancini:2013b} and
\citet{huitson:2013}. The value of the Sun was taken from
\citet{berdyugina:2005}. The error bars have been suppressed for
clarity. Note that some stars appear twice or more. The black dots
refer to the values estimated for the case of HAT-P-36 (this
work). In particular, the lower is the value estimated by taking
the weighted mean from the starpots detected with the Cassini
telescope in April 2013, while the higher value was calculated
from starpots detected with the CA 1.23\,m telescope in April
2014.}%
\label{starspots_tem_plot}
\end{figure*}

The observations of multiple planetary transits across the same
starspot or starspot complex could provide another type of
precious information \citep{sanchis:2011}. Indeed, thanks to the
good alignment between the stellar spin axis and the perpendicular
to the planet's orbital plane, one can measure the shift in
position of the starspot between the transit events and constrain
the alignment between the orbital axis of the planet and the spin
axis of the star with higher precision than from the measurement
of the RM effect (e.g., \citealt{tregloan:2013}).

In our case, we have two consecutive transits observed in 2014
April 14 and 18 and we might wonder if the planet has crossed the
same starspot complex in those transit events. According to Eq.\,1
of \citet{mancini:2014b}, the same starspot can be observed after
consecutive transits or after some orbital cycles, presuming that
in the latter case the star performs one or more rotations around
its axis. Since the projected obliquity measured by the RM effect
(see Sect.\,\ref{sec_4}) indicates spin-orbit alignment, we expect
to find similar values for the starspot parameters in our fits.
Examining Table\,\ref{tab:hatp36_phot_parameters_2}, we note that
they seem to agree on co-latitude, size and contrast within their
1-$\sigma$ errors. Owing to its size, the starspot on April 14
should still be seen four days later (if it is on the visible side
of the star). That means that the starspot should have travelled
$\sim 315^{\circ}$ in four days, giving a rotation period of $\sim
4.57$\,days, which is too fast compared with that at the stellar
equator estimated from the sky-projected rotation rate and the
stellar radius, that is
%
%\begin{linenomath}
\begin{equation}
P_{\mathrm{rot}}\approx \frac{2 \pi
R_{\star}}{v\sin{i_{\star}}}\sin{i_{\star}}= (16.9 \pm
4.3\,\mathrm{d})\sin{i_{\star}}, %
\label{Eq:1}
\end{equation}
%\end{linenomath}
%
where ${i_\star}$ is the inclination of the stellar rotation axis
with respect to the line of sight. If we consider reversing the
rotational direction so that the starspot is travelling right to
left, then the starspot had to travel for $\sim 45^{\circ}$ in
four days; but this implies that the stellar rotation period at
co-latitude of $87^{\circ}$ would be now too large ($\sim
32$\,days) and that HAT-P-36\,b has a retrograde orbit. Since the
latter hypothesis is excluded by the geometry of the RM effect
that we observed (see Sect.\,\ref{sec_4.3}) and since four days are
not sufficient for the starspot to rotate around the back of the
star and then appear on the left hemisphere, we conclude that the
starspot observed on 2014 April 14 is different from those on
April 18.

% Sect. 3.3
%%%%%%%%%%%%%%%%%%%%%%%%%%%%%%%%%%%%%%%%%%%%%%%%%%%%%%%%%%%%%%%%%%%
\subsection{Frequency analysis of the time-series light curves} %
\label{sec_3.3}
%%%%%%%%%%%%%%%%%%%%%%%%%%%%%%%%%%%%%%%%%%%%%%%%%%%%%%%%%%%%%%%%%%%
%
\begin{figure}
\centering \resizebox{\hsize}{!}{\includegraphics{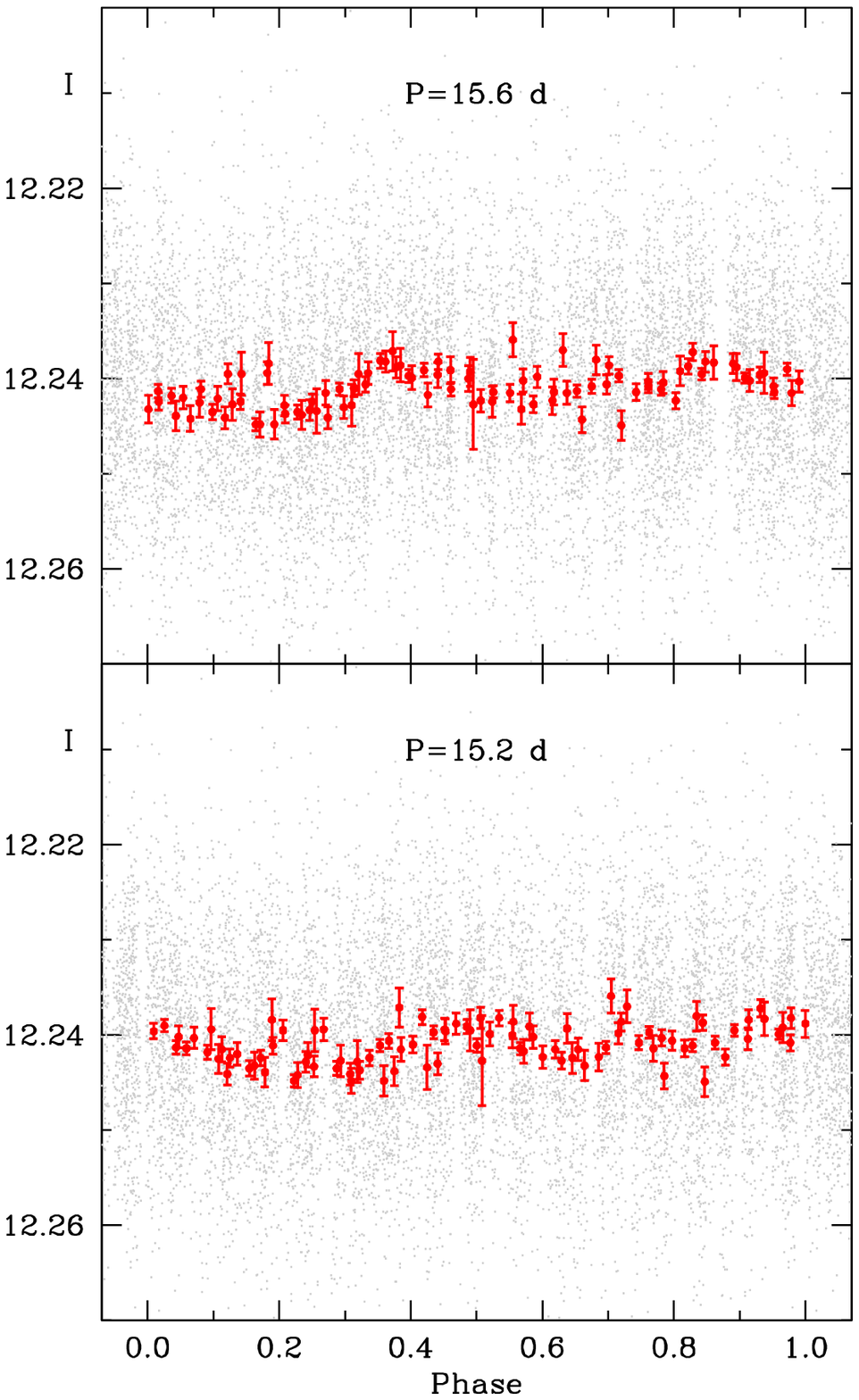}}
\caption{
%{\it Left panel:} Original data (black dots) and average values (red circles with
%errorbars) folded with P=7.787~d, the highest peak in the power spectrum
%of the original data.
{\it Top panel:} Single measurements (grey dots) and average
values on single nights (red circles with errorbars) folded with
$P=15.6$\,d, the best period obtained fitting the data with $f$
and 2$f$ fixed. {\it Bottom panel:} Single measurements (grey
dots) and average values (red circles with errorbars) folded with
$P=15.2$\,d, the best period obtained leaving the frequencies free
to vary. } \label{lc}
\end{figure}
\begin{figure}
\centering \resizebox{\hsize}{!}{\includegraphics{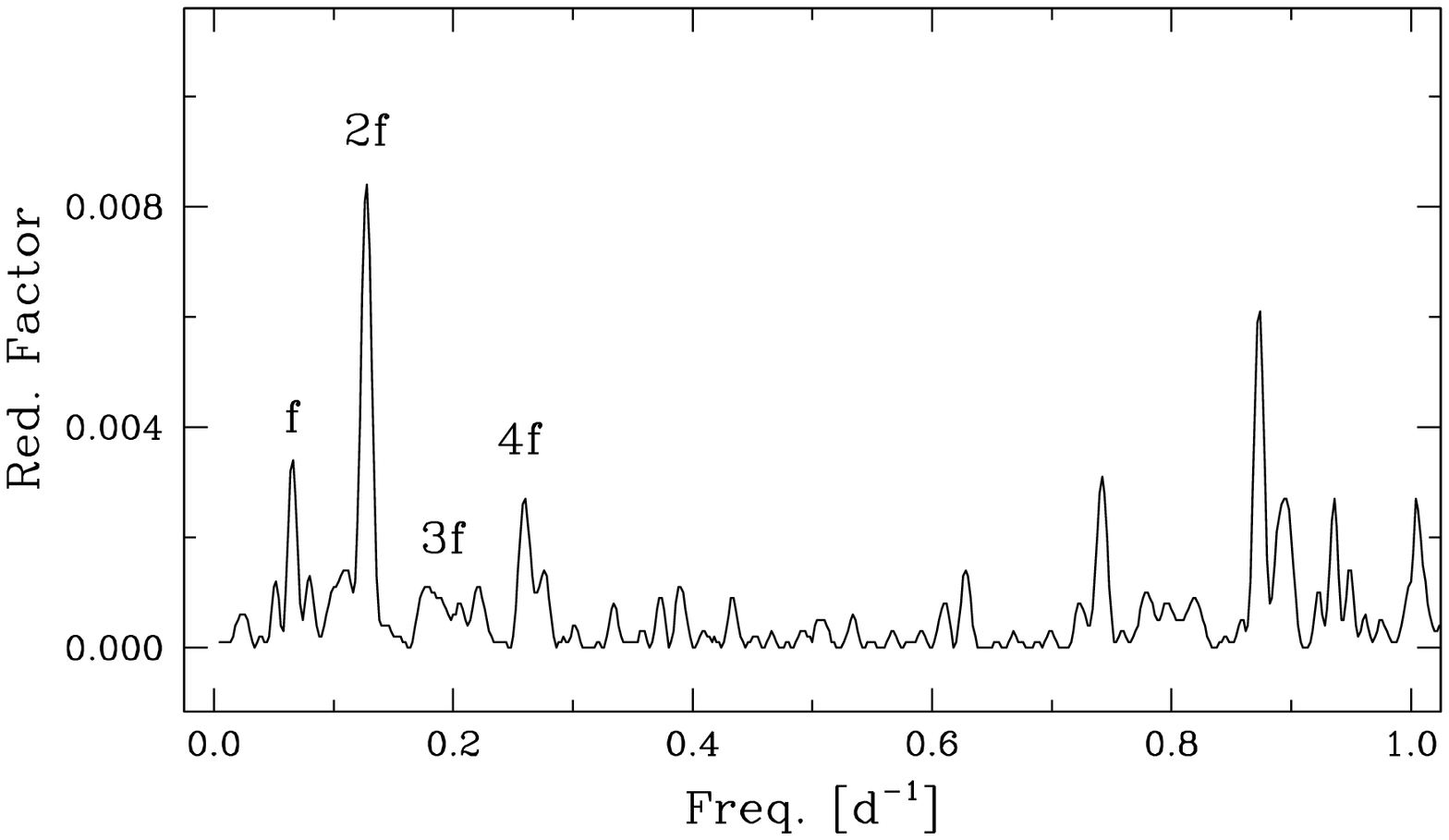}}
\caption{Power spectrum of the original data of HAT-P-36, after
removing the measurements during transit events.} \label{pervani}
\end{figure}

In addition to the new photometric measurements of the transits
presented in this work, dense time-series light curves are
available in the WASP and HAT archives/databases for both the
stars. These data are potentially very useful for detecting the
signatures of the rotational period of each of the two stars. This
is particularly true for the case of HAT-P-36, since the starspots
detected in the photometric light curves (see previous section)
indicate that this star is active. A further confirmation of this
activity is also given by the spectral analysis, as seen in
Sect.\,\ref{sec_4.2}.

We retrieved the photometric time series for HAT-P-36 (SDSS $r$
filter) and WASP-11/HAT-P-10 ($I$ Bessel filter) from the HAT
public archive\footnote{http://hatnet.org}; more specifically, we used
the magnitudes data sets tagged as TF1 \citep{kovacs:2005}. The
WASP data available for WASP-11/HAT-P-10 were downloaded from the
NASA Exoplanet
Archive\footnote{http://exoplanetarchive.ipac.caltech.edu/}. The
in-transit data points were removed from the light curve of each
dataset.

The frequency analysis of the WASP and HAT time-series data
related to WASP-11/HAT-P-10 did not detect any significant
periodicity. Only the HAT data show a non-significant peak at
$f=0.034$~d$^{-1}$. It is close to the synodic month, and the
folded light curve shows a gap in the phase coverage corresponding
to the Full Moon epochs. Therefore, it has been recognized as a
spurious peak due to a combination of instrumental effects and
spectral window.

On the other hand, the analysis of the HAT time-series light curve
of HAT-P-36 (see Fig.\,\ref{lc}) provided more interesting
results. After removing the transit data, the time-series light
curve is composed of 9891 measurements. Both the iterative
sine-wave fitting method \citep[ISWF; ][]{vani} and the
Lomb-Scargle periodogram supplied a lower peak at
$f=0.066$~d$^{-1}$ and a higher one at $2f=0.13$~d$^{-1}$.
Figure\,\ref{pervani} shows the power spectrum obtained with the
ISWF method. The amplitudes of the $2f$ and $f$ components are
$1.40 \pm 0.15$ and $0.88 \pm 0.15$~mmag, respectively. Since the
level of the noise is 0.2~mmag,  both are greater than the $S/N=4.0$
significance threshold \citep{snr4}. Moreover, the harmonics $3f$
and $4f$ are both clearly visible, the latter being more relevant
than the former. HAT-P-36  was measured several times per night
and often over night. Hence we could calculate 97 night averages
(red points in Fig.~\ref{lc}). The frequency analysis of these
averages supplied the same results as the single
measurements.

When the even harmonics have amplitudes larger than those of the
odd ones, the resulting light curve is shaped as a double wave.
Since we know that the star is seen equator-on (see
Sect.\,\ref{sec_4.3}), such a light curve can easily be due to two
groups of starspots that are alternatively visible to the observer. This is an
independent confirmation of the spot model suggested in the
previous subsection. We determined the rotational period in two
ways. First, we searched for the best fit by fixing the frequency
$f$ and its harmonic $2f$. This procedure yields $P_{\rm
rot}=15.2\pm 0.3$~d. Then, we supposed that the differential
rotation and the spread in longitude did not justify this
condition, so we left the two frequency values free to vary,
obtaining $P_{\rm rot}=15.6\pm 0.1$~d. The periods are the same
within the error bars and in excellent agreement with the period
inferred from the stellar parameters, see Eq.\,(\ref{Eq:1}). We
merged the two determinations into $P_{\rm rot}=15.3 \pm 0.4$\,d
to cover both values and respective errorbars. The HAT-P-36 light
curve is shown in Fig.~\ref{lc}, phase-folded with $P_{\rm
rot}=15.6$ (top panel) and $P_{\rm rot}=15.2$ (bottom panel) days.
%The observed full amplitude of the light variability is
%0.0034~mag and the least-squares fit leaves a residual
%\emph{r.m.s.} of 0.0104~mag.

% Sect. 4
%%%%%%%%%%%%%%%%%%%%%%%%%%%%%%%%%%%%%%%%%%%%%%%%%%%%%%%%%%%%%%%%%%%
\section{HARPS-N spectra analysis} %
\label{sec_4}
%%%%%%%%%%%%%%%%%%%%%%%%%%%%%%%%%%%%%%%%%%%%%%%%%%%%%%%%%%%%%%%%%%%

% Sect. 4.1
%%%%%%%%%%%%%%%%%%%%%%%%%%%%%%%%%%%%%%%%%%%%%%%%%%%%%%%%%%%%%%%%%%%
\subsection{Spectroscopic determination of stellar atmospheric parameters}
%\label{SubSec:star-atmo-par}
\label{sec_4.1}
%%%%%%%%%%%%%%%%%%%%%%%%%%%%%%%%%%%%%%%%%%%%%%%%%%%%%%%%%%%%%%%%%%%

The effective temperature ($T_{\rm eff}$), surface gravity
($\log{g_{\star}}$), iron abundance ([Fe/H]) and microturbulence
velocity ($v_{\mathrm{mic}}$) of the two host stars were derived
by using the spectral analysis package MOOG
(\citealp{sneden:1973}, version 2013), and the equivalent widths
(EWs) of iron lines, as described in detail in
\citet{biazzo:2012}. The EWs were measured on the mean spectrum
obtained by averaging all the HARPS-N spectra available for each
of the two stars. Here, $T_{\rm eff}$ was determined by imposing the
condition that the \ion{Fe}{i} abundance does not depend on the
excitation potential of the lines, the microturbulence velocity,
by requiring that the \ion{Fe}{i} abundance is independent on the
line EWs, and $\log{g_{\star}}$ by the \ion{Fe}{i}/\ion{Fe}{ii}
ionization equilibrium condition. The results are reported in
Tables\,\ref{tab:hatp36_final_parameters} and
\ref{tab:wasp11_final_parameters}, for HAT-P-36 and WASP-11,
respectively. The projected rotational velocity was estimated by
means of spectral synthesis analysis with MOOG, yielding
$v\sin{i_{\star}}$ values of $3.0\pm0.5$ km\,s$^{-1}$ and
$0.9\pm0.5$ km\,s$^{-1}$ for HAT-P-36 and WASP-11, respectively,
which are consistent, within the errors, with those obtained from
the RM effect modelling (see Sect.\,\ref{sec_4.3}).

% Sect. 4.2
%%%%%%%%%%%%%%%%%%%%%%%%%%%%%%%%%%%%%%%%%%%%%%%%%%%%%%%%%%%%%%%%%%%
\subsection{Stellar activity index}
%\label{SubSec:star-atmo-par}
\label{sec_4.2}
%%%%%%%%%%%%%%%%%%%%%%%%%%%%%%%%%%%%%%%%%%%%%%%%%%%%%%%%%%%%%%%%%%%
The median values of the stellar activity index estimated from the
HARPS-N spectra turned out to be
$\log{R_{\mathrm{HK}}^{\prime}}=-4.636 \pm0.066$\,dex for HAT-P-36
and $-4.848 \pm 0.029$\,dex for WASP-11/HAT-P-10, indicating a
moderate and low activity, respectively
\citep[e.g.][]{Noyes:1984}. These values agree with
what we have found in Sect.\,\ref{sec_3.3} by analysing the big
photometric data sets from the HATNet amd WASP surveys. The above
values were obtained by adopting $B-V=0.719$ and $0.989$,
respectively, based on the stellar effective temperature and the
color-$T_{\mathrm{eff}}$ conversion table by
\citet{casagrande:2010}. \citet{knutson:2010} found $\log{R_{\mathrm{HK}}^{\prime}}=-4.823$\,dex for
WASP-11/HAT-P-10 
(estimated considering $B-V=1.01$), which is in good agreement with
our value.

Thanks to the activity index, we can also assess the expected
rotation period from the level of the stellar activity, obtaining
for HAT-P-36 $P_{\mathrm{rot}}=18.3 \pm 3.1$\,d and $18.7 \pm
2.9$\,d using \citet{Noyes:1984} and \citet{mamajek:2008}
calibration scales, respectively, in quite good agreement,
within the observational uncertainties, with what was
estimated in Sect.\,\ref{sec_3.3}. For WASP-11/HAT-P-10, we
obtained $40.8 \pm 1.8$\,d and $40.9 \pm 2.2$\,d, respectively.

Finally, we can get a clue on stellar age by applying the
activity-age calibration proposed by \citet{mamajek:2008},
obtaining $2$\,Gyr and $5$\,Gyr for HAT-P-36 and WASP-11/HAT-P-10,
respectively. However, we notice that the ages derived from
stellar rotation and activity could be altered in the TEP systems
because of star-planet tidal interactions, and any discrepancy
with the values derived from the isochrones (see
Sect.\,\ref{sec_5}) may be due to this.

% Sect. 4.3
%%%%%%%%%%%%%%%%%%%%%%%%%%%%%%%%%%%%%%%%%%%%%%%%%%%%%%%%%%%%%%%%%%%
\subsection{Determination of the spin-orbit alignment} %
\label{sec_4.3}
%%%%%%%%%%%%%%%%%%%%%%%%%%%%%%%%%%%%%%%%%%%%%%%%%%%%%%%%%%%%%%%%%%%

The analysis of the photometric data allowed us a direct
measurement, without reliance on theoretical stellar models, of
the stellar mean density, $\rho_{\star}$
\citep{seager:2003,sozzetti:2007} and, in combination with the
spectroscopic orbital solution, the planetary surface gravity,
$g_{\mathrm{p}}$ \citep{southworth:2007}, for each system.
Exploiting the measured values of $\rho_{\star}$,
$T_{\mathrm{eff}}$ and [Fe/H], we made use of the Yonsei-Yale
evolutionary tracks \citep{demarque:2004} to determine stellar
characteristics, including $M_{\star}$ and $R_{\star}$. The
HARPS-N RV data sets were then fitted using a model that accounts
both for the RV orbital trend and the RM anomaly. 

We used the RM model that was elaborated and described in \citet{covino:2013} and \citet{esposito:2014},
and we used the same least-square minimization algorithm to adjust it to the data. We set as
free the barycentric radial velocity, $\gamma$, the mass of the
planet, $M_{\mathrm{p}}$, the projected stellar rotational
velocity, $v\,\sin{i_{\star}}$, the projected spin-orbit angle,
$\lambda$ and the linear limb-darkening coefficient $u$. All the
other relevant parameters were kept fixed to the values determined
from the spectroscopic and light curves analysis.
The best-fitting RV models are illustrated in Fig.\,\ref{hatp36RM}
and \ref{wasp11RM}, superimposed on the datasets. In particular,
the sky-projected spin-orbit misalignment angle was
$\lambda= -14 \pm 18$\,deg for HAT-P-36 and $\lambda=7 \pm 5$\,deg
for WASP-11/HAT-P-10, indicating an alignment of the stellar spin
with the orbit of the planet for both the systems. The
final error bars were computed by a bootstrapping approach. The
greater uncertainty in HAT-P-36 is of course because
the transit was not completely observed. The reduced
$\chi^{2}_{\nu}$ maps in the $\lambda - v \sin{i_{\star}}$ plane
do not point out any clear correlation between $\lambda$ and
$v\,\sin{i_{\star}}$. 

The precise knowledge of the rotational period of a star, allows
estimating the stellar spin inclination angle, $i_{\star}$, and
thus the true spin-orbit alignment angle, $\psi$. This is exactly
the case for HAT-P-36, for which the analysis of the long
time-series data set recorded by HATNet survey allowed us to
determine the stellar rotational period. By using $P_{\rm
rot}=15.3 \pm 0.4$\,d (Sect.\,\ref{sec_3.3}), we estimated that
$i_{\star}=65^{\circ} \pm 34^{\circ}$. Then, we used Eq.\,(7) in
\citet{winn:2007},
%
%\begin{linenomath}
\begin{equation}
\cos{\psi}=\cos{i_{\star}}\cos{i}+\sin{i_{\star}}\sin{i}\cos{\lambda}, %
\label{Eq:2}
\end{equation}
%\end{linenomath}
%
to derive the true misalignment angle, obtaining
$\psi=25^{+36}_{-25}$. The major source of error in the
determination of both $i_{\star}$ and $\psi$, estimated through the propagation 
of uncertainties, comes from the large
relative error in the measurement of $v\sin{i_{\star}}$; the
observation of a full transit would have reduced the errors by a
factor of 2 or more.

%%%%%%%%%%%%%%%%%%%%%%%%%%%%%%%%%%%%%%%%%%%%%%%%%%%%%%%
%% Figure: chi^2 maps
%%%%%%%%%%%%%%%%%%%%%%%%%%%%%%%%%%%%%%%%%%%%%%%%%%%%%%%
%\begin{figure}
%\centering
%\includegraphics[width=8.2cm]{chi2_maps.eps}
%\caption{The reduced $\chi^{2}_{\nu}$ map in the $\lambda - v
%\sin{i_{\star}}$ parameter space for HAT-P-36 (upper panel) and
%WASP-11/HAT-P-10 (lower panel) planetary systems. The numerical
%values of $\chi_{\nu}^{2}$ on the contours are reported.}%
%\label{chi2maps}
%\end{figure}
%%%%%%%%%%%%%%%%%%%%%%%%%%%%%%%%%%%%%%%%%%%%%%%%%%%%%%

% Sect. 5
%%%%%%%%%%%%%%%%%%%%%%%%%%%%%%%%%%%%%%%%%%%%%%%%%%%%%%%%%%%%%%%%%%%
\section{Physical parameters of the two systems}
\label{sec_5}
%%%%%%%%%%%%%%%%%%%%%%%%%%%%%%%%%%%%%%%%%%%%%%%%%%%%%%%%%%%%%%%%%%%

For calculating the full physical properties of each planetary
system, we used the HSTEP methodology (see
\citealp{southworth:2012}, and references therein). The values of
the photometric parameters $r_{\rm A}+r_{\rm b}$, $k$ and $i$ were
first combined into weighted means. The orbital eccentricity was
fixed to zero. We added in the measured $P$ from the light curve
analysis, $T_{\mathrm{eff}}$ and [Fe/H] from the spectroscopic
analysis, and the velocity semi-amplitude of the star, $K_{\rm
A}$, measured from the RVs. For $K_{\rm A}$ we used the values
from \citet{bakos:2012} and \citet{bakos:2009}. These come from
spectra with reasonable coverage of the orbits of the host stars,
so are more precise than our own values which come from data
obtained only during or close to transit. We estimated a starting
value for the velocity semi-amplitude of the planet, $K_{\rm p}$,
and used this to determine a provisional set of physical
properties of the system using standard formulae (e.g.,
\citealp{hilditch:2001}).

We then interpolated within a set of tabulated predictions from a
theoretical stellar model to find the expected stellar radius and
$T_{\mathrm{eff}}$ for our provisional mass and the observed
[Fe/H]. The value of $K_{\rm p}$ was then iteratively refined to
maximise the agreement between the observed and predicted
$T_{\mathrm{eff}}$, and the provisional $(R_{\rm A}+R_{\rm b})/a$
and measured $r_{\rm A}+r_{\rm b}$ values. This $K_{\rm b}$ was
then used alongside the other quantities given above to determine
the physical properties of the system. This process was performed
over all possible ages for the stars, from the zero-age to the
terminal-age main sequence, and the overall best-fitting age and
resulting physical parameters found.

The above process was performed using each of five different sets
of theoretical models: Claret \citep{claret:2004}, Y$^2$
\citep{demarque:2004}, BaSTI \citep{pietrinferni:2004}, VRSS
\citep{vandenberg:2006} and DSEP \citep{dotter:2008}. The final
result of this analysis was five sets of properties for each
system, one from each of the five sources of theoretical models
(see Tables\,\ref{tab:hatp36:model} and \ref{tab:wasp11:model}).
For the final value of each output parameter, we calculated the
unweighted mean of the five estimates from using the different
sets of model predictions. Statistical errorbars were propagated
from the errorbars in the values of all input parameters.
Systematic errorbars were assigned based on the inter-agreement
between the results from the five different stellar models.

Tables \ref{tab:hatp36_final_parameters} and
\ref{tab:wasp11_final_parameters} give the physical properties we
found for HAT-P-36 and WASP-11, respectively. Our results are in
good agreement, but they are more precise than previous determinations.

%%%%%%%%%%%%%%%%%%%%%%%%%%%%%%%%%%%%%%%%%%%%%%%%%%%%%%
% Table -- HAT-P-36 final parameters
%%%%%%%%%%%%%%%%%%%%%%%%%%%%%%%%%%%%%%%%%%%%%%%%%%%%%%
\begin{table*}
\tiny
\centering %
\caption{Physical parameters of the planetary system HAT-P-36 derived in this work.} %
\label{tab:hatp36_final_parameters} %
\begin{tabular}{l c c c c}
\hline %
\hline  \\[-8pt]%%
Parameter & Nomen. & Unit & This Work & \citet{bakos:2012}\\
\hline  \\[-6pt]%%
\multicolumn{1}{l}{\textbf{Stellar parameters}} \\
Spectral class                \dotfill &                      &              & G5\,V                  & $...             $ \\ %
Effective temperature         \dotfill & $T_{\mathrm{eff}}$   & K            & $5620   \pm 40     $   & $5560   \pm 100  $ \\ %
Metal abundance               \dotfill & [Fe/H]               &              & $+0.25  \pm 0.09   $   & $+0.26  \pm 0.10 $ \\ %
Projected rotational velocity \dotfill & $v\,\sin{i_{\star}}$ & km\,s$^{-1}$ & $3.12   \pm 0.75   $   & $3.58   \pm 0.5  $ \\ %
Rotational period              \dotfill & $P_{\mathrm{rot}}$  & days & $15.3 \pm 0.4 $   & $...$ \\ %
Linear LD coefficient         \dotfill & $u$                  &              & $0.81   \pm 0.17   $   & $...$               \\ [3pt]%
Mass                          \dotfill & $M_{\star}$       & $M_{\sun}$ & $1.030  \pm 0.029 \pm 0.030 $ & $1.022  \pm 0.049$ \\ %
Radius                        \dotfill & $R_{\star}$       & $R_{\sun}$ & $1.041  \pm 0.013 \pm 0.010 $ & $1.096  \pm 0.056$ \\ %
Mean density                  \dotfill & $\rho_{\star}$ & $\rho_{\sun}$ & $0.913  \pm 0.027  $ & $...$                        \\ %
Logarithmic surface gravity   \dotfill & $\log{g_{\star}}$    & cgs          & $4.416  \pm 0.010 \pm 0.004 $ & $4.37   \pm 0.04$ \\ %
Age                           \dotfill &                      & Gyr          & $4.5_{-1.4\,-2.8}^{+2.4\,+3.1} $ & $6.6_{-1.8}^{+2.9}$ \\ %
\hline \\[-6pt]%
\multicolumn{1}{l}{\textbf{Planetary parameters}} \\
%Stellar fractional radius     \dotfill & $r_{\star}$        & ...                & $0.2026  \pm 0.0024 $  & ...                 \\ %
%Planetary fractional radius   \dotfill & $r_{\mathrm{p}}$   & ...                & $0.02609 \pm 0.00083$  & ...                 \\ %
Mass              \dotfill & $M_{\mathrm{p}}$    & $M_{\mathrm{Jup}}$    & $1.852  \pm 0.088 \pm 0.036 $ & $1.832  \pm 0.099 $ \\ %
Radius            \dotfill & $R_{\mathrm{p}}$    & $R_{\mathrm{Jup}}$    & $1.304  \pm 0.021 \pm 0.013 $ & $1.264  \pm 0.071 $ \\ %
Mean density      \dotfill & $\rho_{\mathrm{p}}$ & $\rho_{\mathrm{Jup}}$ & $0.737  \pm 0.095 \pm 0.007 $ & $0.84   \pm 0.14 $ \\ %
Surface gravity   \dotfill & $g_{\mathrm{p}}$    & m\,s$^{-2}$           & $27.0   \pm 1.4    $ & $28     \pm 3             $ \\ %
Equilibrium temperature \dotfill & $T_{\mathrm{eq}}$ & K & $1788 \pm 15$ & $1823 \pm 55$  \\ %
Safronov number             \dotfill & $\Theta$       &         & $0.0658 \pm  0.0030 \pm 0.0006$  & $0.067 \pm 0.005$  \\ %
\hline \\[-6pt]%
\multicolumn{1}{l}{\textbf{Orbital parameters}} \\
Time of mid-transit         \dotfill & $T_{0}$        & BJD(TDB)& $2\,455\,565.18167 \pm 0.00036 $ & $2\,455\,565.18144 \pm 0.00020 $ \\ %
Period              \dotfill & $P_{\mathrm{orb}}$            & days    & $1.32734683 \pm 0.00000048 $ & $1.327347 \pm 0.000003 $ \\ %
Semi-major axis     \dotfill & $a$            & au      & $0.02388 \pm 0.00022 \pm 0.00023$ & $0.0238 \pm 0.0004$ \\ %
Inclination         \dotfill & $i$            & degree  & $85.86   \pm 0.21   $ & $86.0   \pm 1.3   $ \\  [3pt] %
RV-curve semi-amplitude       \dotfill & $K_{\rm A}$          &  m\,s$^{-1}$ & $316    \pm 39$ $^{a}$ & $334.7  \pm 14.5 $ \\ %
Barycentric RV                \dotfill & $\gamma$             & km\,s$^{-1}$ & $-16.327\pm 0.006  $   & $-16.29 \pm 0.10 $ \\  [3pt] %
%Eccentricity        \dotfill & $e$  & ...     & $0.0460  \pm 0.0047 $ & $0.063  \pm 0.032 $ \\ %
%Argument of Periastron      \dotfill & $\omega$       & degree  & $80.83   \pm 3.73   $ & $95     \pm 63    $ \\ %
Projected spin-orbit angle  \dotfill & $\lambda$      & degree  & $-14     \pm 18     $ & $...              $ \\ %
True spin-orbit angle  \dotfill & $\psi$      & degree  & $ 25_{-25}^{+38}            $ & $...              $ \\ %

\hline %
\end{tabular}
\tablefoot{Where two errorbars are given, the first refers to the
statistical uncertainties and the second to the systematic errors.
\\ $^{a}$ This value of $K_{\mathrm{A}}$ was determined from
our RV data, which were taken during transit time only. The value
reported by \citet{bakos:2009}, which is based on RV data with a
much better coverage of the orbital phase, was therefore preferred
for the determination of the other physical parameters of the
system (see Sect.\ref{sec_5}).}
\end{table*}

%%%%%%%%%%%%%%%%%%%%%%%%%%%%%%%%%%%%%%%%%%%%%%%%%%%%%%
% Table -- WASP-11 final parameters
%%%%%%%%%%%%%%%%%%%%%%%%%%%%%%%%%%%%%%%%%%%%%%%%%%%%%%

\begin{table*}
\tiny
\centering %
\caption{Physical parameters of the planetary system WASP-11/HAT-P-10 derived in this work.} %
\label{tab:wasp11_final_parameters} %
\begin{tabular}{l c c c c c}
\hline \\[-8pt]%%
Parameter & Nomen. / Unit & This Work & \citet{west:2009} & \citet{bakos:2009} & \citet{wang:2014} \\
\hline \hline  \\[-6pt]%
\multicolumn{1}{l}{\textbf{Stellar parameters}} \\
Spectral Class             \dotfill & & K3\,V                & K & K & $...$   \\ %
Effective temperature      \dotfill & $T_{\mathrm{eff}}$ (K) & $4900   \pm 65     $ & $4800  \pm 100   $ & $4980  \pm 60   $ & $... $ \\ %
Metal abundance            \dotfill & $\mathrm{[Fe/H]}$      & $+0.12  \pm 0.09   $ & $+0.0  \pm 0.2   $ & $+0.13 \pm 0.08 $ & $... $ \\ %
Proj. rotational velocity  \dotfill & $v\sin{i_{\star}}$ (km\,s$^{-1})$  & $1.04    \pm 0.15    $ & $<6.0 $ & $0.5 \pm 0.2$ & $...$ \\ %
Linear LD coefficient      \dotfill & $u$                & $0.78   \pm 0.16   $ & $...$ & $...$ & $...$                 \\ [3pt]%
Mass                       \dotfill & $M_{\star}$ ($M_{\sun}$) & $0.806  \pm 0.038 \pm 0.013  $ & $0.77^{+0.10}_{-0.08} $ & $0.83 \pm 0.03$ & $0.862 \pm 0.014$ \\ %
Radius                     \dotfill & $R_{\star}$ ($R_{\sun}$) & $0.772  \pm 0.014 \pm 0.004  $ & $0.74^{+0.04}_{-0.03} $ & $0.79 \pm 0.02$ & $0.784^{+0.018}_{-0.011}$ \\ %
Mean density               \dotfill & $\rho_{\star}$ ($\rho_{\sun}$) & $1.748  \pm 0.074  $ & $...$ & $...$ & $1.789^{+0.072}_{-0.116}$ \\ %
Logarithmic surface gravity\dotfill & $\log{g_{\star}}$ (cgs) & $4.569  \pm 0.018 \pm 0.002  $ & $4.45   \pm 0.20  $ & $4.56 \pm 0.02 $ & $... $ \\ %
Age                        \dotfill & (Gyr)             & $7.6_{-3.0\,-1.8}^{+5.8\,+1.6}   $ & $...$ & $7.9 \pm 3.8$ & $...$ \\ %
\hline \\[-6pt]%
\multicolumn{1}{l}{\textbf{Planetary parameters}} \\
Mass                    \dotfill & $M_{\mathrm{p}}$ ($M_{\mathrm{Jup}}$)       & $0.492  \pm 0.023 \pm 0.005 $ & $0.53 \pm 0.07 $ & $0.487 \pm 0.018 $ & $0.526 \pm 0.019 $\\ %
Radius                  \dotfill & $R_{\mathrm{p}}$ ($R_{\mathrm{Jup}}$)       & $0.990  \pm 0.022 \pm 0.005 $ & $0.91^{+0.06}_{-0.03}$ & $1.005^{+0.032}_{-0.027}$ & $0.999^{+0.029}_{-0.018}$ \\ %
Mean density            \dotfill & $\rho_{\mathrm{p}}$ ($\rho_{\mathrm{Jup}}$) & $0.475  \pm 0.026 \pm 0.002 $ & $0.69^{+0.07}_{-0.11}$ & $0.594 \pm 0.052$ & $0.526^{+0.035}_{-0.046}$ \\ %
Surface gravity         \dotfill & $g_{\mathrm{p}}$ (m\,s$^{-2}$)              & $12.45  \pm 0.50 $ & $14.45^{+1.66}_{-1.33}$& $12.02 \pm 0.83 $ & $...$ \\ %
Equilibrium temperature \dotfill & $T_{\mathrm{eq}}$ (K)                       & $992    \pm 14$    & $960 \pm 70$ & $1020 \pm 17$ & $1006.5^{+16.4}_{-14.6}$ \\ %
Safronov number         \dotfill & $\Theta$                                    & $0.0539 \pm 0.0019 \pm 0.0003$ & $...$ & $0.053 \pm 0.002$ & $...$ \\ %
\hline \\[-6pt]%
\multicolumn{1}{l}{\textbf{Orbital parameters}} \\
\vspace{0.1cm} %
Time of mid-transit \dotfill & $T_{0}$ (BJD$_{\mathrm{TDB}}$) &
\begin{tabular}{c}
$2\,454\,729.90716$ \\
$\pm 0.00020$       \\
\end{tabular} &
\begin{tabular}{c}
$2\,454\,473.05588$ \\
$\pm 0.00020$       \\
\end{tabular} &
\begin{tabular}{c}
$2\,454\,759.68683$ \\
$\pm 0.00016$       \\
\end{tabular} &
\begin{tabular}{c}
$2\,454\,808.07904$ \\
$\pm 0.00012$       \\
\end{tabular} \\
\vspace{0.1cm} %
Period \dotfill & $P_{\mathrm{orb}}$ (days)    &
\begin{tabular}{c}
$3.72247967$         \\
$\pm 0.00000045$     \\
\end{tabular} &
\begin{tabular}{c}
$3.722465$         \\
$\pm 0.000007$     \\
\end{tabular} &
\begin{tabular}{c}
$3.7224747$         \\
$\pm 0.0000065$     \\
\end{tabular} &
\begin{tabular}{c}
$3.72247669$         \\
$\pm 0.00000181$     \\
\end{tabular}      \\ %
Semi-major axis  \dotfill & $a$ (au)      &
\begin{tabular}{c}
$0.04375$ \\
$\pm 0.00070 \pm 0.00023$
\end{tabular}
& $0.043  \pm 0.002$ & $0.0435 \pm 0.0006$ & $0.04473 \pm 0.00024$ \\ %
\vspace{0.1cm} %
Inclination      \dotfill & $i$ (degree)  & $89.03  \pm 0.34   $ & $89.8_{-0.8}^{+0.2} $ &$88.6^{+0.5}_{-0.4}$ & $89.138^{+0.503}_{-0.470}$ \\ [3pt] %
RV curve semi-amplitude    \dotfill & $K_{\rm A}$ (m\,s$^{-1}$) & $82.7   \pm 4.2$ $^{a}$   & $82.1 \pm 7.4     $ & $74.5 \pm 1.8     $ & $76.16_{-2.58}^{+2.67}  $ \\ %
Barycentric RV             \dotfill & $\gamma$ (km\,s$^{-1}$)  & $4.8951 \pm 0.0021 $ & $4.9077 \pm 0.0015 $ & $3.95 \pm 0.43$ & $...$ \\ [3pt] %
Proj. spin-orbit angle \dotfill & $\lambda$ (degree)  & $7     \pm 5      $ & $...$ & $...$ & $...$ \\ %
\hline %
\end{tabular}
\tablefoot{Where two errorbars are given, the first refers to the
statistical uncertainties and the second to the systematic
errors.\\
$^{a}$ This value of $K_{\mathrm{A}}$ was determined from our RV
data, which were taken during transit time only. The value
reported by \citet{bakos:2009}, which is based on RV data with a
much better coverage of the orbital phase, was therefore preferred
for the determination of the other physical parameters of the
system (see Sect.\ref{sec_5}).}
\end{table*}

% Sect. 6
%%%%%%%%%%%%%%%%%%%%%%%%%%%%%%%%%%%%%%%%%%%%%%%%%%%%%%%%%%%%%%%%%%%
\section{Discussion}
\label{sec_6}
%%%%%%%%%%%%%%%%%%%%%%%%%%%%%%%%%%%%%%%%%%%%%%%%%%%%%%%%%%%%%%%%%%%

%%%%%%%%%%%%%%%%%%%%%%%%%%%%%%%%%%%%%%%%%%%%%%%%%%%%%%%
%% Figure: Lambda plot
%%%%%%%%%%%%%%%%%%%%%%%%%%%%%%%%%%%%%%%%%%%%%%%%%%%%%%%
\begin{figure*}
\centering
\includegraphics[width=18.4cm]{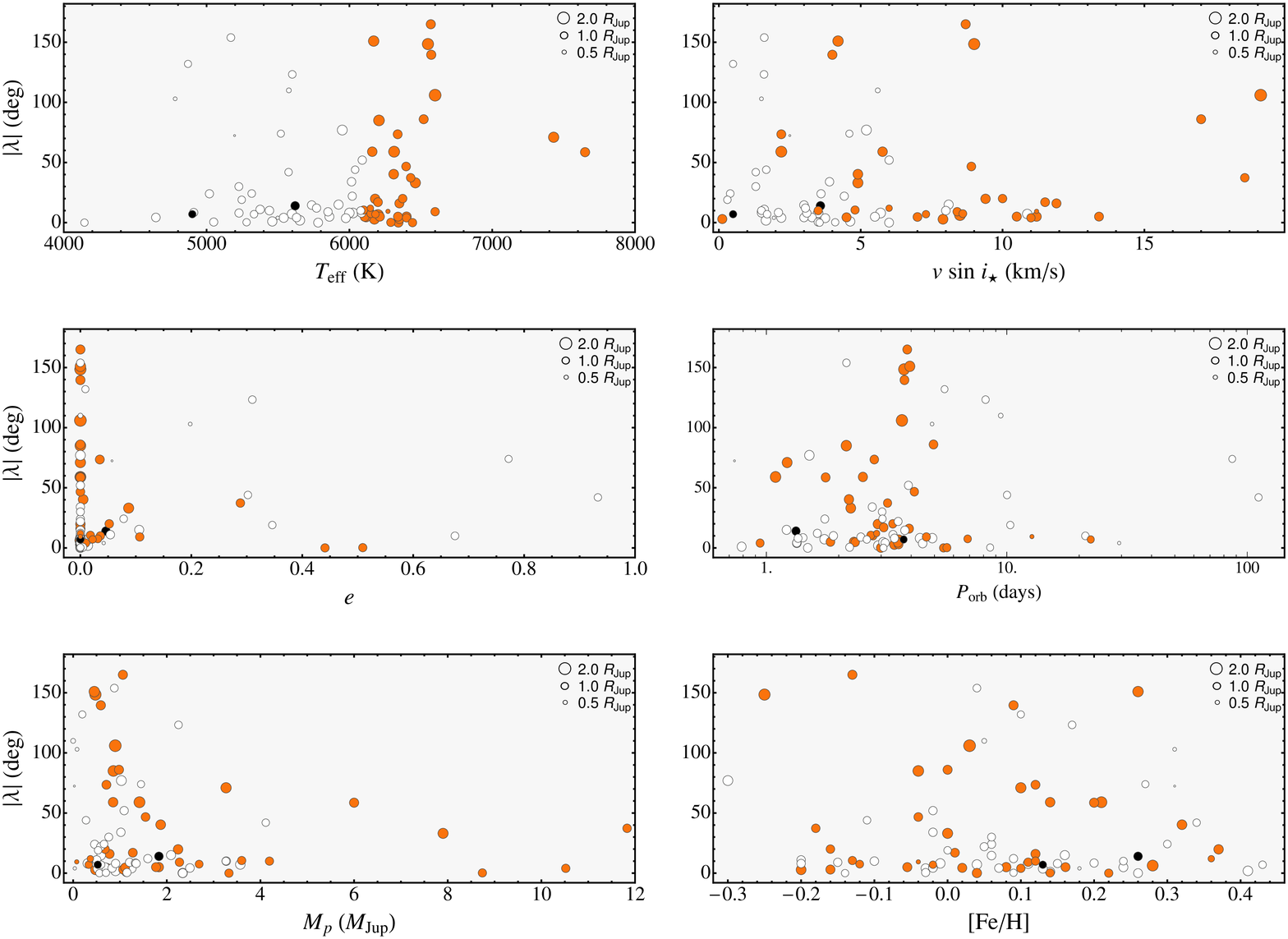}
\caption{Sky-projected orbital obliquity as a function of several
stellar, orbital and planetary parameters. Orange (white) circles
indicate systems in which the parent stars has an effective
temperature higher (lower) than 6100\,K \citep{dawson:2014}. The
size of each circle is proportional to the corresponding planetary
radius. The data have been taken from TEPcat
\citep{southworth:2011}. The black points indicate the two TEP
systems examined in this work, HAT-P-36 and WASP-11/HAT-P-10. The
error bars have been suppressed for clarity.
%The inset in panel (e) zooms on
%$0\,M_{\mathrm{Jup}} < M_{\mathrm{p}} < 4\,M_{\mathrm{Jup}}$.
} \label{rm_plots}
\end{figure*}

As introduced in Sect.\,\ref{sec_1}, orbital obliquity could be an
important parameter to determine in order to trace the physical
processes that happened during migration phase of current hot
Jupiters. Several empirical trends have been presented to
corroborate this hypothesis. TEP systems, in which the parent star
is relatively cool ($T_{\mathrm{eff}} \lesssim 6100$\,K), should
be much more aligned than those with hotter stars, because the
realignment of cool star's convective envelopes can occur on
timescales shorter than planet's orbital decay
\citep{winn:2010,albrecht:2012b,dawson:2014}. The larger
spin-orbit misalignment expected for the hot stars (F-type stars
with $T_{\mathrm{eff}} > 6100-6200$\,K) could be also explained by
considering they have a smaller convective zone than that of the
cool stars. This implies that the tidal dissipation is much less
efficient for the hot stars than the cold ones
\citep{lai:2012,valsecchi:2014}.

It was also noted that TEP systems in which the hot Jupiter is
massive ($M_{\mathrm{p}} \gtrsim 3\,M_{\mathrm{Jup}}$) tend to
have lower spin-orbit angles, the parent star being much more
affected by planet's tidal influence \citep{hebrard:2011}.
However, several theoretical studies were not able to explain the
above correlations (e.g., \citealp{rogers:2013,xue:2014}) and, as
stressed by \citet{esposito:2014}, there are planetary systems
composed by cool stars that are highly misaligned.

With this work, we contribute to enlarge the sample, adding the
measurement of the sky-projected orbital obliquity for other two
TEP systems. They are HAT-P-36 and WASP-11/HAT-P10, both composed
by stars with $T_{\mathrm{eff}} < 6100$\,K, whose spin results to
be well aligned with the planetary-orbit axis. They are reported
in Fig.\,\ref{rm_plots} (black circles), together with the other
82 known TEP systems\footnote{The data were taken from TEPcat
\citep{southworth:2011}. The brown dwarfs KELT-1 and WASP-30 were
excluded from the sample.}, as a function of stellar effective
temperature, projected rotational velocity, eccentricity, orbital
period, planetary mass and stellar metallicity. Following
\citet{dawson:2014}, we divided the data in two groups, based on
the parent stars temperature; they are shown with orange circles
if $T_{\mathrm{eff}}>6100$\,K and with empty circles for the
opposite case. An inspection of these plots does not allow to
highlight any clear trend, but, on the contrary, the values of
$\lambda$ are quite randomly distributed in the various parameter
spaces, suggesting that the observed diversity of stellar
obliquities may be a consequence of more complicated interactions
with outer planets.

We also plotted in Fig.\,\ref{rm_ar_plots} the projected orbital
obliquity of the cool-star systems, i.e., following
\citet{anderson:2015}, those with $T_{\mathrm{eff}}<6150$\,K, as a
function of orbital distance in units of stellar radii,
$a/R_{\star}$. However, we do not see an absolute confinement of
$\lambda$ at any particular orbital-separation range.

According to TEPCat, there are now 12 TEP systems for which we
have the measurement of the true obliquity, i.e. the angle,
$\psi$, between the axes of the stellar rotation and the planetary
orbit. These values are plotted against the stellar temperature in
Fig.\,\ref{psi_plot} that, again, does not highlight any
correspondence between the two quantities.

\begin{figure}
\centering
\includegraphics[width=9.cm]{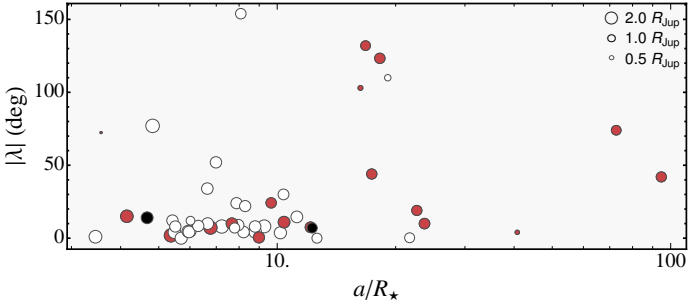}
\caption{The projected orbital obliquity as a function of scaled
orbital distance, $a/R_{\star}$, for those systems with
$T_{\mathrm{eff}}<6150$\,K \citep{anderson:2015}. The empty
circles represent near-circular orbits ($e< 0.1$ or consistent
with zero), while the light-red circles depict eccentric orbits.
The size of each circle is proportional to the corresponding
planetary radius. The data have been taken from TEPcat
\citep{southworth:2011}. The black points indicate the two TEP
systems examined in this work, HAT-P-36 and WASP-11/HAT-P-10. The
error bars have been suppressed
for clarity.} %
\label{rm_ar_plots}
\end{figure}

Based on a recent gyrochronology work by \citet{meibom:2015}, the
rotation period of HAT-P-36, estimated in Sect.\,\ref{sec_3.3},
should correspond to an age of about 1.8 Gyr, which is close to
the lower limit reported in
Table\,\ref{tab:hatp36_final_parameters}, as derived from our
isochrone-fitting analysis. Since the planet is very close to its
host, we may speculate about tidal interactions affecting the
evolution of the orbit and of the stellar spin. Assuming a
modified tidal quality factor $Q^{\prime}_{*} = 10^{6}$, the
remaining lifetime of the system -- i.e., the time left before the
planet plunges into the star -- is only  35 Myr
\citep{metzger:2012}. Such a rapid orbital decay should be
accompanied by a negative transit time variation O-C of about 20
seconds in ten years that would be easy to measure with a
space-borne photometer. On the other hand, if we assume
$Q^{\prime}_{*} \sim 10^{7}-10^{8}$, as suggested by
\citet{ogilvie:2007}, the orbital decay and alignment timescales
are of the order of a few hundred Myr or a few Gyr, respectively.
They are shorter than or comparable to the estimated age of the
star. In conclusion, in this scenario the present alignment could
result from the tidal interaction during the main-sequence
lifetime, while the remaining lifetime of the system is at least
of a few hundred Myr. In any case, HAT-P-36 is an interesting
system to study the orbital and stellar spin evolution, as
resulting from the tidal interaction and the magnetic wind braking
of the star.

\begin{figure*}
\centering
\includegraphics[width=18.4cm]{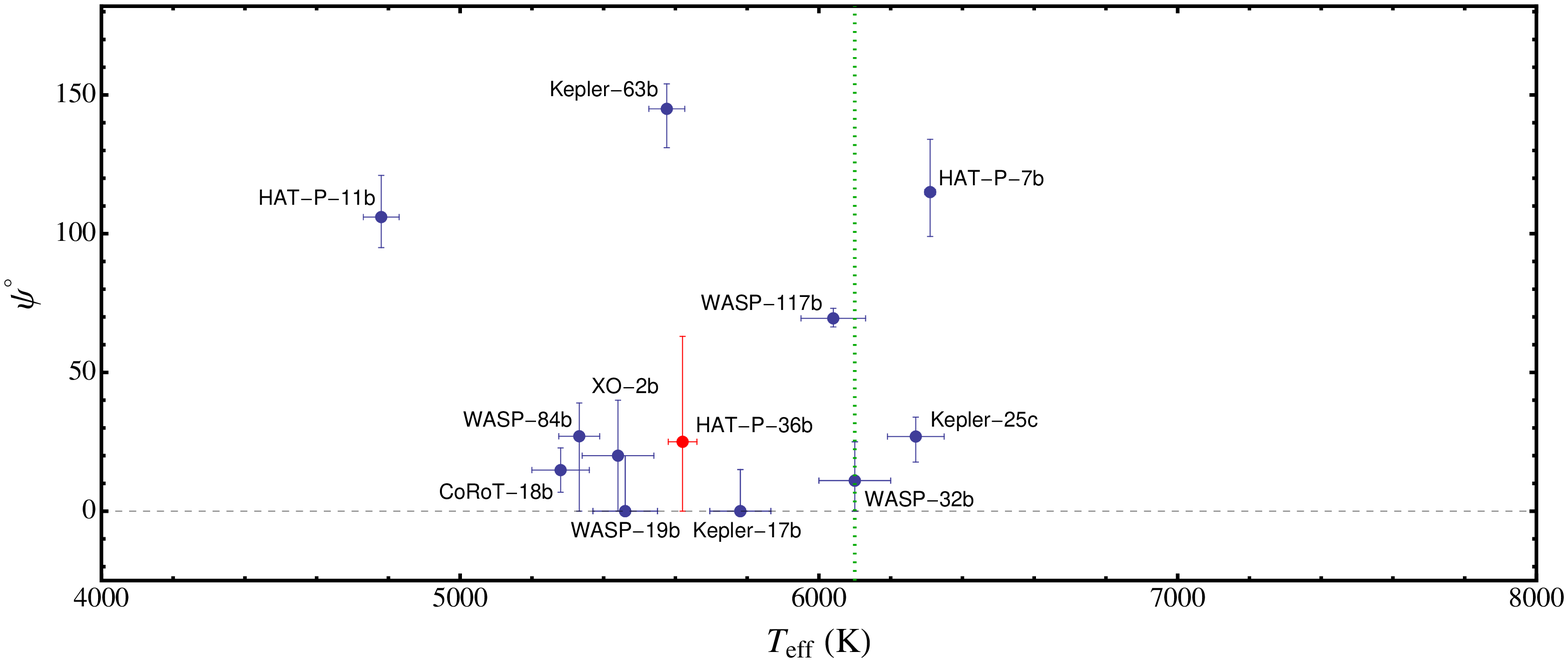}
\caption{True orbital obliquity as a function of the effective
stellar temperature for 12 TEP systems. Data taken from TEPCat.
The red point indicates the location of HAT-P-36 (this work). The
temperature range is the same as that in the upper-left panel of
Fig.\,\ref{rm_plots}. The green-dotted line separates the cool TEP
systems from those with $T_{\mathrm{eff}}>6100$\,K
\citep{dawson:2014}.} \label{psi_plot}
\end{figure*}

For WASP-11, the relatively large semi-major axis and the small
mass of the planet imply, on the other hand, a weak tidal
interaction with an angular momentum exchange timescale that is
comparable or longer than the age of the star. Therefore, the
observed projected alignment of the system is likely to be of
primordial origin.

% Sect. 7
%%%%%%%%%%%%%%%%%%%%%%%%%%%%%%%%%%%%%%%%%%%%%%%%%%%%%%%%%%%%%%%%%%%
\section{Summary and conclusions}
\label{sec_7}
%%%%%%%%%%%%%%%%%%%%%%%%%%%%%%%%%%%%%%%%%%%%%%%%%%%%%%%%%%%%%%%%%%%

In the framework of the \texttt{GAPS} programme, we are observing
a sample of TEP systems  hosting close-in hot Jupiters, with the
HARPS-N spectrograph. The goal is to measure their spin-orbit
alignment (by observing the RM effect) and study if this quantity
is correlated with other physical properties. Here, we have
presented an exhaustive study of the HAT-P-36 and WASP-11/HAT-P10
planetary systems, based on data collected with four telescopes
during transit events.

By analysing these new photometric and spectroscopic data, we
revise the physical parameters of these two systems, finding that
previous determinations agree with our more accurate results, within the
uncertainties.

Interestingly, we observed anomalies in three photometric transit
light curves of HAT-P-36 that, after appropriate modelling,
turned out to be compatible with starspot complexes on the
photosphere of the star. The main characteristics of the starspots
were estimated and compared with those observed in other TEP
systems. The HAT-P-36 starspot activity is confirmed by the
analysis of the HARPS-N spectra, which give a stellar activity
index of $\log{R_{\mathrm{HK}}^{\prime}}=-4.65$\,dex, and by
study of the whole photometric data set collected by the HATNet
survey. The frequency analysis of this time-series data revealed a
clear modulation in the light curve caused by starspots, allowing
us to get an accurate measurement of the rotational period of the
star, $P_{\mathrm{rot}}=15.3 \pm 0.4$\,days. A similar study,
performed for WASP-11/HAT-P-10 on both the WASP and HATNet survey
light curves, did not highlight any clear photometric modulation.

The RM effect was partially covered in HAT-P-36 (because of bad
weather conditions) and wholly observed in WASP-11/HAT-P-10.
Thanks to the high-resolution HARPS-N spectra, the sky projection
of the orbital obliquity was successfully measured for both the
systems, indicating a good spin-orbit alignment. In particular,
for the HAT-P-36 system, we were able to also estimate its real
obliquity, obtaining $\psi=25^{+38}_{-25}$\,degrees. Our results
are thus in agreement with the idea that stars with relatively
cool photospheres should have small spin-orbit misalignment
angles. However, looking at the entire sample of known TEP
systems, for which we have an accurate estimation of $\lambda$,
there are several exceptions to this scenario. In this context, the case of HAT-P-18\,b 
\citep{esposito:2014} is emblematic.
The collection of more data is thus mandatory for disentangling the
issue and understanding whether orbital obliquity really holds imprints of
past migration processes that dramatically affected the evolution
of giant planets.

%  Acknowledgements
%%%%%%%%%%%%%%%%%%%%%%%%%%%%%%%%%%%%%%%%%%%%%%%%%%%%%%%%%%%%%%%%%%%
\begin{acknowledgements}
%This paper is based on observations collected with the following
%telescopes: 3.58\,m Telescopio Nazionale Galileo (TNG), operated
%on the island of La Palma (Spain) by the Fundaci\'{o}n Galileo
%Galilei of the INAF (Istituto Nazionale di Astrofisica) at the
%Spanish Observatorio del Roque de los Muchachos of the Instituto
%de Astrofísica de Canarias, in the frame of the programme Global
%Architecture of Planetary Systems (GAPS); Zeiss 1.23\,m telescope
%at the Centro Astron\'{o}mico Hispano Alem\'{a}n (CAHA) in Calar
%Alto (Spain); Cassini 1.52\,m telescope at the Astronomical
%Observatory of Bologna in Loiano (Italy); IAC\,80\,cm telescope at
%the Teide Observatory on the island of Tenerife (Spain), operated
%by the Instituto de Astrof\'{i}sica de Canarias.
The HARPS-N instrument has been built by the HARPS-N Consortium, a
collaboration between the Geneva Observatory (PI Institute), the
Harvard-Smithonian Center for Astrophysics, the University of St.
Andrews, the University of Edinburgh, the Queen's University of
Belfast, and INAF. Operations at the Calar Alto telescopes are
jointly performed by the Max-Planck Institut f\"{u}r Astronomie
(MPIA) and the Instituto de Astrof\'{i}sica de Andaluc\'{i}a
(CSIC). The reduced light curves presented in this work will be
made available at the CDS (http://cdsweb.u-strasbg.fr/). The GAPS
project in Italy acknowledges support from INAF through the
``Progetti Premiali'' funding scheme of the Italian Ministry of
Education, University, and Research. We acknowledge the use of the
following internet-based resources: the ESO Digitized Sky Survey;
the TEPCat catalog; the SIMBAD data base operated at CDS,
Strasbourg, France; and the arXiv scientific paper preprint
service operated by Cornell University.
\end{acknowledgements}

\bibliographystyle{aa}

\begin{appendix}

%(e.g, \citealp{southworth:2012a,southworth:2012b,%
%mancini:2013a,mancini:2014b,covino:2013,ciceri:2013,esposito:2014}),

%%%%%%%%%%%%%%%%%%%%%%%%%%%%%%%%%%%%%%%%%%%%%%%%%%%%%
\section{Supplementary tables}
\label{Appendix_A}
%%%%%%%%%%%%%%%%%%%%%%%%%%%%%%%%%%%%%%%%%%%%%%%%%%%%%
The tables in this Appendix contain the detailed results of the
predictions of different sets of stellar evolutionary models for
the HAT-P-36 and WASP-11/HAT-P-10 planetary systems. The final
values of the physical parameters for each of the two systems (see
Tables\,\ref{tab:hatp36_final_parameters} and
\ref{tab:wasp11_final_parameters}), are calculated by taking the
unweighted mean of the five estimates of the different sets of
model predictions (see Sect\,\ref{sec_5}).

\begin{table*}%
 %\tiny %
 %\flushleft %
 \caption{\label{tab:hatp36:model} Derived physical properties of the HAT-P-36 system based on the prediction of different theoretical models.}
 \begin{tabular}
 {l r@{\,$\pm$\,}l r@{\,$\pm$\,}l r@{\,$\pm$\,}l r@{\,$\pm$\,}l r@{\,$\pm$\,}l r@{\,$\pm$\,}l}
 \hline \hline\\[-8pt]%
% \ & \mc{This work} & \mc{This work} & \mc{This work} & \mc{This work} & \mc{This work}  \\
 \ & \mc{({\sf Claret} models)} & \mc{({\sf Y$^2$} models)} & \mc{({\sf BaSTI} models)} & \mc{({\sf VRSS} models)} & \mc{({\sf DSEP} models)} \\
 \hline\\[-6pt]%
$K_{\rm b}$     (\kms) & 196.19   &   4.95    & 196.09   &   1.79   & 193.00   &   1.62    & 193.00   &   2.76    & 196.17   &   0.32   \\
$M_{\star}$    (\Msun) & 1.0503   & 0.0840    & 1.0486   & 0.0287   & 1.0000   & 0.0259    & 1.0000   & 0.0426    & 1.0500   & 0.0045   \\
$R_{\star}$    (\Rsun) & 1.048    & 0.029     & 1.047    & 0.012    & 1.031    & 0.013     & 1.031    & 0.018     & 1.048    & 0.010    \\
$\log g_{\star}$ (cgs) & 4.4192   & 0.0132    & 4.4190   & 0.0100   & 4.4121   & 0.0092    & 4.4121   & 0.0109    & 4.4192   & 0.0085   \\ [2pt] %
$M_{\rm p}$    (\Mjup) & 1.877    & 0.130     & 1.875    & 0.088    & 1.816    & 0.087     & 1.816    & 0.095     & 1.876    & 0.079    \\
$R_{\rm p}$    (\Rjup) & 1.313    & 0.038     & 1.312    & 0.020    & 1.291    & 0.021     & 1.291    & 0.025     & 1.313    & 0.018    \\
$\rho_{\rm p}$ (\pjup) & 0.776    & 0.049     & 0.776    & 0.046    & 0.789    & 0.047     & 0.789    & 0.048     & 0.776    & 0.047    \\
\safronov\             & 0.0654   & 0.0033    & 0.0654   & 0.0030   & 0.0665   & 0.0030    & 0.0665   & 0.0032    & 0.0654   & 0.0030   \\ [2pt] %
$a$               (AU) & 0.024040 & 0.000605  & 0.024028 & 0.000219 & 0.023651 & 0.000198  & 0.023651 & 0.000338  & 0.024038 & 0.000036 \\
Age              (Gyr) & \erc{4.3}{2.4}{1.4}  & \erc{4.0}{1.1}{0.8} & \erc{7.6}{4.3}{1.9}  & \erc{5.0}{1.5}{2.0}  & \erc{1.7}{7.5}{0.1} \\
 \hline
 \end{tabular}
 \tablefoot{In each case $g_{\rm p} = 27.0 \pm 1.4$\mss,
 $\rho_{\star} = 0.913 \pm 0.027$\psun\ and $T_{\mathrm{eq}} = 1788 \pm 15$\,K.}
\end{table*}

\begin{table*} %
 %\tiny %
 \flushleft %
 \caption{\label{tab:wasp11:model} Derived physical
 properties of the WASP-11/HAT-P-10 system based on the prediction of different theoretical models.}
 \begin{tabular}{l r@{\,$\pm$\,}l r@{\,$\pm$\,}l r@{\,$\pm$\,}l r@{\,$\pm$\,}l r@{\,$\pm$\,}l }
 \hline \hline\\[-8pt]%
% \ & \mc{This work} & \mc{This work} & \mc{This work} & \mc{This work} & \mc{This work}  \\
 \ & \mc{({\sf Claret} models)} & \mc{({\sf Y$^2$} models)} & \mc{({\sf BaSTI} models)} & \mc{({\sf VRSS} models)} & \mc{({\sf DSEP} models)} \\
 \hline\\[-6pt]%
$K_{\rm b}$     (\kms) & 129.9   &   3.2     & 127.7   &   2.0     & 127.5   &   1.9      & 127.5   &   2.3      & 128.4   &   2.1      \\
$M_{\star}$    (\Msun) & 0.846   & 0.063     & 0.804   & 0.039     & 0.800   & 0.036      & 0.800   & 0.043      & 0.818   & 0.038      \\
$R_{\star}$    (\Rsun) & 0.785   & 0.023     & 0.772   & 0.013     & 0.771   & 0.015      & 0.771   & 0.014      & 0.777   & 0.012      \\
$\log g_{\star}$ (cgs) & 4.576   & 0.016     & 4.569   & 0.016     & 4.568   & 0.015      & 4.568   & 0.018      & 4.571   & 0.018      \\ [2pt] %
$M_{\rm p}$    (\Mjup) & 0.508   & 0.028     & 0.492   & 0.020     & 0.490   & 0.019      & 0.490   & 0.024      & 0.497   & 0.023      \\
$R_{\rm p}$    (\Rjup) & 1.006   & 0.031     & 0.989   & 0.019     & 0.987   & 0.020      & 0.987   & 0.023      & 0.995   & 0.022      \\
$\rho_{\rm p}$ (\pjup) & 0.467   & 0.028     & 0.475   & 0.025     & 0.476   & 0.026      & 0.476   & 0.027      & 0.472   & 0.026      \\
\safronov\             & 0.0531  & 0.0021    & 0.0540  & 0.0016    & 0.0541  & 0.0017     & 0.0541  & 0.0019     & 0.0537  & 0.0019     \\ [2pt] %
$a$               (AU) & 0.04447 & 0.00111   & 0.04373 & 0.00069   & 0.04365 & 0.00065    & 0.04365 & 0.00080    & 0.04398 & 0.00070    \\
\end{tabular}
\begin{tabular}{cccccccccc}
 Age              (Gyr) & \erc{~~~~~~~~~~~  5.7}{5.8}{4.8} & \erc{~~~~~~~~~~~~~~~  8.9}{3.9}{3.9} & ~~~~~~~~~~~~~~~~... & ~~~~~~~~~~~~~~~~~~~~... & \erc{~~~~~~~~~~~~~~~~~~~6.2}{2.7}{2.5}   \\
 \hline \end{tabular}
\tablefoot{In each case $g_{\rm p} = 12.45 \pm 0.50$\mss,
$\rho_{\star} = 1.748 \pm 0.074$\psun\ and $T_{\mathrm{eq}} = 992
\pm 14$\,K. We were not able to constrain a reasonable value for
the age of the WASP-11/HAT-P-10 system by using the BaSTI and VRSS
models, because the evolution of a star of the mass of WASP-11 is
small for times shorter than the age of the Galaxy.}
\end{table*}

\end{appendix}

\end{document}